\documentclass[twocolumn,
showpacs,preprintnumbers,prd,amsmath,amssymb,aps,floats,floatfix]{revtex4}
\usepackage{graphicx}
\usepackage{dcolumn}
\usepackage{bm}
\def\nnd{\end{document}}

\def\nnb{\nonumber}
\def\de{\delta}

\def\be{\begin{equation}}
\def\ee{\end{equation}}
\def\mn{\mu\nu}

\newcommand{\bea}{\begin{eqnarray}}
\newcommand{\eea}{\end{eqnarray}}
\newcommand{\beaa}{\begin{equation}\begin{array}{rl}}
\newcommand{\eeaa}{\end{array}\end{equation}}
\newcommand{\bwt}{\begin{widetext}}
\newcommand{\ewt}{\end{widetext}}

\def\bbra#1{\Bigg \{ #1  \Bigg \} }

\def\cma{\,,}
\def\hA{\widehat A}

\def\hWp{\widehat W^{+}}
\def\bWp{\overline W^{+}}
\def\hWm{\widehat W^{-}}
\def\bWm{\overline W^{-}}
\def\hZ{\widehat Z}
\def\m_z{m_{textrm{Z}}}
\def\bZ{\overline Z}

\def\wsep{ \nnb \\ &&}

\def\ssep{\right. \nnb\\ && \left.}
\def\eed{\end{document}}

\def\lx{{\stackrel{\leftharpoonup}{X}}}
\def\rx{{\stackrel{\rightharpoonup}{X}}}
\def\al{\alpha}

\def\al{\alpha}
\def\be{\beta}

\def\eb{\eea\bea}
\def\al{{\alpha}}
\def\my#1{ #1 }
\def\nkf#1{ #1 }

\def\sbkf#1{\big ( #1 \big )}

\def\sbbkf#1{\bigg ( #1 \bigg )}
\def\mbbkf#1{\bigg [ #1 \bigg ]}

\makeatother

\begin{document}

\title{ Constraints on the electroweak chiral Lagrangian from the precision data }
\author{ Sukanta Dutta $^{a,\,b}$}
\author{ Kaoru Hagiwara $^{b,\,c}$}
\author{Qi-Shu Yan $^{d,\,e}$\,\, }
\author{Kentaroh Yoshida $^f$\,\, }
\affiliation{$^a$ SGTB Khalsa College, University of
  Delhi, Delhi-110007 India}
\author{  }
\affiliation{$^b$ KEK Theory Division,
Tsukuba,         305-0801 Japan}
\affiliation{$^c$ The Graduate University for Advanced Studies (SOKENDAI), 
Tsukuba, 305-0801 Japan}
\affiliation{$^d$  Department of Physics, National Tsing Hua University,
Hsinchu, Taiwan }
\affiliation{$^e$ National Center of Theoretical Sciences (Theory Division),
101, Section 2 Kuang Fu Road, Hsinchu, Taiwan}
\affiliation{$^f$ Kavli Institute for Theoretical Physics, University of California,
Santa Barbara, CA. 93106-4030, U.S.}



\begin{abstract}
In the framework of the effective field theory method,
we use the experimental data 
and the perturbative unitarity bounds to determine the values and  uncertainty
of all the $11$ chiral coefficients ($\al_i,\, i=0,\, \cdots,\, 10$)
of the standard electroweak chiral Lagrangian. 
Up to linear terms in $\al_i$, we provide the one-loop renormalization 
group equations of all the chiral coefficients, which are calculated in
the Feynman-'t Hooft gauge using the modified minimal subtraction scheme. 
With the improved renormalization group equations to sum over the
logarithmic corrections, 
we analyze the current experimental uncertainty of oblique 
correction parameters, $S(\Lambda)$ and $T(\Lambda)$. 
We find that, due to the large uncertainty in 
the triple gauge-boson coupling measurements, the parameter space of positive
$S(\Lambda)$ for $\Lambda > 1$ TeV is still allowed by 
the current experimental data. 
$T(\Lambda)$ tends to increase with $\Lambda$ even in the presence
of the operators that contribute to the triple and quartic
gauge-boson couplings.

\end{abstract}
\pacs{11.10.Gh, 11.10.Hi, 11.15.Ex, 10.30.Rd, 12.15.Ji, 12.15.Lk}
\maketitle

\section{Introduction}
Electroweak symmetry breaking (EWSB) mechanism is the 
most important issue which will be explored 
at the Large Hadronic Collider (LHC). 
TeV-scale supersymmetric theories  suggests that 
the electroweak symmetry is spontaneously broken
by fundamental Higgs fields.
On the contrary, QCD-like theories do not have fundamental
scalar fields and suggest that the electroweak symmetry is 
dynamically broken by fermion-pair condensates \cite{ws,hill}.

In order to obtain hints on the TeV scale physics that leads to the EWSB, 
it is useful to examine the consequence of 
the electroweak precision data of the bosonic
sector in an integrated fashion. 
In this paper, we study the scale dependence of the
coefficients of the electroweak chiral Lagrangian,
and examine carefully if the present precision data
exclude models with $S(\Lambda \sim 1 \, \textrm{TeV)} < 0$
conclusively.  We pay particular attention to the
magnitude and sign of the coefficients of the
operators that contribute only to the triple and
quartic gauge-boson couplings and hence weakly
constrained by the present data, because they
enter in the $\beta$ functions of the running $S$ and $T$ parameters.
Part of our findings have been reported in Ref. \cite{our}.

We follow the standard analysis of the chiral 
Lagrangian method \cite{Gasser:1983yg, Gasser:1984gg, Harada:2003jx} and 
include $11$ operators up to mass dimension 
four in the electroweak chiral Lagrangian (EWCL) 
\cite{Appelquist:1980ae,Longhitano:1980iz,Longhitano:1980tm,Appelquist:1993ka}. 
This study extends the work of 
Bagger {\it et. al.} \cite{Bagger},
who considered the effects of those operators
that contribute to the weak boson two point
functions only, {\it i.e.} $3$ out of the $11$ operators.
We consider all the $11$ operators, among which
$5$ operators contribute to the triple gauge boson couplings (TGCs)
and $9$ contribute to the quartic gauge boson couplings (QGCs).
We consider constraints from the TGC  measurements
at LEP2 \cite{Heister:2001qt,Abbiendi:2003mk,
Achard:2004ji,Schael:2004tq} and at the TeVatron \cite{Abazov:2005ys} as well as 
those from the perturbative unitarity
to bound the QGCs.

In the framework of the effective field theory 
method \cite{georgi}, all the $11$ couplings  
in the EWCL are renormalization scale $\mu $ dependent quantities.
Hence the electroweak precision data constrain the 
chiral coefficients at the low scale, $\mu=m_Z$. 
In order to find the connection between $S(m_Z)$ and
$S(\Lambda)$, we extend the previous one-loop RGE \cite{Herrero:1993nc} of the EWCL 
used by Bagger {\it et. al.} \cite{Bagger}.
Along with the previous studies, we assume the   validity of 
the perturbation theory and the absence of any 
additional resonances between $m_Z$ and $\Lambda$. The  new $\beta$ functions takes account of all the 
the $11$ chiral coefficients. The improved RGE makes it possible to analyze
the effects of those operators which contribute only to the
three-point and four-point gauge boson couplings on
the uncertainty of $S(\Lambda)$ and $T(\Lambda)$ 
at the scale $\Lambda$ of new physics.

By utilizing the RGE to sum over the logarithmic
corrections of quantum fluctuations and 
by analyzing the current experimental uncertainty 
of TGC and QGC,
we find that in the most conservative perturbative 
calculation, the central value of $S(1\,\, \textrm{TeV})$ and its corresponding 
$1\, \sigma$ error  reads as  
\begin{equation}
\begin{array}{rl}
S(1\,\, \textrm{TeV}) = &-0.02\pm 0.20\,.
\end{array}
\label{st1tev}
\end{equation}
We observe that 
the current electroweak precision data, especially the 
data from the TGC 
measurements, have not reached the
precision to fix the signs of $S(\Lambda)$  above  $\Lambda >  1 \,\textrm{TeV}$. 
If we remove the
effects of the chiral coefficients that contribute
only to TGC and QGC,
the central value and its $1\,\sigma$ error  reads as 
\bea
S(1 \,\,\textrm{TeV}) = &-0.14\pm 0.09\,,
\eea
reproducing the result of Ref. \cite{Bagger}.

In our analysis, we will use the formalism of nonlinear realization of EWSB, {\it i.e.} the gauged
nonlinear $\sigma$ model.  
We consider the set of all the dimension 4 operators
which are even under the transformation of discrete symmetries, C and P.
We omit all dimension $6$ or higher order operators.
There are works, for instance Ref. \cite{Barbieri:2004qk}, which attempt
to constrain the electroweak 
symmetry breaking models by including some of dimension $6$ operators. 
By including more operators one can only make the allowed area of the  
$S(\Lambda)-T(\Lambda)$ plane larger, and hence our
conclusion that $S(\Lambda=1\textrm{TeV}) > 0$ is still allowed
persists.

This paper is organized as follows.
In Sec. II, the bosonic sector of the EWCL 
in our analysis is introduced.
In Sec. III, we briefly review the constraints on 
the chiral coefficients from the electroweak precision data
on the gauge boson two point functions (A),
those on the TGC (B), and those from the perturbative unitarity
of the weak boson scattering amplitudes (C).
In Sec. IV, we list the improved renormalization group
equations (RGEs). 
In Sec. V, we study the experimental 
uncertainty of $S(\Lambda)$-$T(\Lambda)$ for $\Lambda=0.3 \textrm{TeV}$,
$1 \textrm{TeV}$, and $3 \textrm{TeV}$ by using the improved RGEs.
We close the paper in Sec. VI with discussions and conclusions. 
Two appendices are added to introduce our method to calculate
the $\beta$ functions of the chiral coefficients (A) and the treatment
of the ghost terms (B).

\section{ The operators in our analysis }
In our analysis,  
we consider the following $14$ bosonic operators
which preserve discrete symmetries, $C$ and $P$ 
\cite{Appelquist:1980ae,Longhitano:1980iz,Longhitano:1980tm,Appelquist:1993ka}:
\bea
{\cal L}_{EW} &=&- \frac{1}{g^2} {\bar H_1} - \frac{1}{{g'}^2}{\bar H_2} 
- v^2 {\bar {\cal L}_{W/Z}} + \al_0 v^2 {\bar {\cal L}_0} \nnb \\
&& + \sum_{i=1}^{10} \al_i {\bar {\cal L}_i} \,, \label{ewl} \\
{\bar H_1}&=&\frac{1 }{2 } tr(W_{\mn} W^{\mn}) \label{h1} \cma \\
{\bar H_2}&=&\frac{1 }{4 } B_{\mn} B^{\mn} \label{h2} \cma\\
{\bar {\cal L}_{W/Z}}&=&\frac{1 }{ 4} tr(V_{\mu} V^{\mu}) \label{lwz} \cma
\eea
where $g$ and ${g'}$ are the gauge couplings of $SU(2)_L$ and $U(1)_Y$
gauge groups, respectively. The Nambu-Goldstone bosons are parameterized
in the nonlinear form as
\bea
U      &=& \exp\left\{ i\,\frac{2 \pi^a T^a}{v} \right\}\,.
\eea
The gauge covariant derivative, local gauge fields and their gauge covariant
field strength are given as
\bea
V_{\mu}&=&(\partial_{\mu} U)  U^{\dagger} + i W_{\mu}  - i U B^Y_{\mu} U^{\dagger} \,. \label{dif} \\
W_{\mu}&=&W_{\mu}^a T^a\,\\
B^Y_{\mu}&=&B_{\mu} T^3\,,\\
W_{\mu\nu}&=&\partial_{\mu} W_{\nu}-\partial_{\nu} W_{\mu}+ i [W_{\mu},W_{\nu}]\,,\\
B_{\mu\nu}&=&\partial_{\mu} B_{\nu}-\partial_{\nu} B_{\mu}\,,
\eea
where $T^a=\tau^a/2$, and $\tau^a$ are the Pauli matrices.

The couplings $\al_i$ form the $11$ dimensional 
parameter space of the EWCL, where the corresponding operators ${\cal L}_i$ are given as
\bea
{\bar {\cal L}_0}&=&  \frac{1 }{ 4}[tr({\cal T} V_\mu)] [tr({\cal T} V^\mu)]\label{l0} \cma \\
{\bar {\cal L}_1}&=& \, \frac{1}{ 2} B_{\mu\nu}tr({\cal T} W^{\mu\nu})\cma \\
{\bar {\cal L}_2}&=&i \frac{1 }{ 2} B_{\mu\nu}tr({\cal T} [V^\mu,V^\nu])\cma \\
{\bar {\cal L}_3}&=&i        \,\, \,    tr( W_{\mu\nu}[V^\mu,V^\nu])\cma \\
{\bar {\cal L}_4}&=&  \,     \,\, \,      [tr(V_\mu V_\nu)]^2\cma \\
{\bar {\cal L}_5}&=&  \,     \,\, \,      [tr(V_\mu V^\mu)]^2\cma \\
{\bar {\cal L}_6}&=&  \,     \,\, \,     tr(V_\mu V_\nu)tr({\cal T} V^\mu)tr({\cal T} V^\nu)\cma \\
{\bar {\cal L}_7}&=&  \,      \,\,\,     tr(V_\mu V^\mu)[tr({\cal T} V^\nu)]^2\cma \\
{\bar {\cal L}_8}&=&  \,   \frac{1}{ 4} [tr({\cal T} W_{\mu\nu})]^2\cma \\
{\bar {\cal L}_9}&=&i  \frac{1}{ 2} tr({\cal T} W_{\mu\nu})tr({\cal T} [V^\mu,V^\nu])\cma \\
{\bar {\cal L}_{10}}&=& \, \frac{1}{ 2} [tr({\cal T} V_\mu)tr({\cal T} V_\nu)]^2\,,
\label{ewcle}
\eea
with
\bea
{\cal T}&\equiv& 2 U T^3 U^{\dagger}\,.
\eea
Each operator in the Lagrangian ${\cal L}_{EW}$
is invariant under the following local $SU(2)_L\times U(1)_Y$ 
gauge transformation
\begin{equation}
\begin{array}{rcl}
U            &\rightarrow& g_L U g_Y^\dagger \cma \\
W_\mu        &\rightarrow& g_L  W_\mu g_L^\dagger- i g_L \partial_\mu g_L^\dagger \cma \\
W_{\mu\nu}   &\rightarrow& g_L W_{\mu\nu} g_L^\dagger \cma  \\
B^Y_\mu      &\rightarrow& B^Y_\mu - i g_Y \partial_\mu g_Y^\dagger \cma  \\
B_{\mu\nu}   &\rightarrow& B_{\mu\nu} \cma \\
V_{\mu}      &\rightarrow& g_L V_{\mu} g_L^\dagger \cma \\
{\cal T}     &\rightarrow& g_L {\cal T} g_L^\dagger \cma
\end{array}
\label{eq:gaugetransformation}
\end{equation}
where the gauge transformation factors $g_L$ and $g_Y $ are defined as
\begin{equation}
\begin{array}{rcl}
g_L (x)&\equiv& \exp{\bbra{- i g {\alpha^{a}}(x) T^a } } \,,\\
g_Y (x)&\equiv& \exp{\bbra{- i g' \beta(x)       T^3 } } \,.
\end{array}
\end{equation}
Here $\alpha^a(x)$ and $\beta(x)$ are the real parameters of 
the gauge transformation of $SU(2)_L$ and $U(1)_Y$, respectively.

While operators ${\bar {\cal L}_{W/Z}}$ and ${\bar {\cal L}_0}$
contribute to the vector boson mass term, 
the operators ${\bar {\cal L}_i},\, i=1,\,\cdots,\,10$
contribute to their   kinetic terms, TGC and QGC,
as tabulated in Table \ref{vertex}.

\begin{table}
\begin{ruledtabular}
\begin{tabular}{| c | c | c | c | }
 & $2$ pt. vtx. & 3 pt. vtx. (TGC) & 4 pt. vtx. (QGC) \\ \hline
$v^2 {\bar {\cal L}_0 }$         &   $\surd$       &           &  \\
\hline
${\bar {\cal L}_1}$          &   $\surd$       & $\surd$   &  \\
\hline
${\bar {\cal L}_2}$          &                 & $\surd$   &  $\surd$ \\
\hline
${\bar {\cal L}_3}$          &                 & $\surd$   &  $\surd$ \\
\hline
${\bar {\cal L}_4}$          &                 &           &  $\surd$ \\
\hline
${\bar {\cal L}_5}$          &                 &           &  $\surd$ \\
\hline
${\bar {\cal L}_6}$          &                 &           &  $\surd$ \\
\hline
${\bar {\cal L}_7}$          &                 &           &  $\surd$ \\
\hline
${\bar {\cal L}_8}$          &   $\surd$       & $\surd$   &  $\surd$ \\
\hline
${\bar {\cal L}_9}$          &                 & $\surd$   &  $\surd$ \\
\hline
${\bar {\cal L}_{10}}$       &                 &           &  $\surd$ \\
\end{tabular}
\end{ruledtabular}
\caption{ Operators and their contributions to $2$, $3$, and $4$ point
gauge boson vertices. }
\label{vertex}
\end{table}

Accordingly we can classify the chiral coefficients
into three groups: {\bf (1)} $\al_0$, $\al_1$, and $\al_8$
contribute to the weak boson two-point functions, and are
constrained by the electroweak precision data \cite{pdg2006}; 
{\bf (2)} $\al_2$, $\al_3$, and $\al_9$ contribute to the
three-point couplings but not to the two-point functions,
and are constrained by the TGC measurements \cite{Heister:2001qt,Abbiendi:2003mk,
Achard:2004ji,Schael:2004tq}; 
{\bf (3)} $\al_4$, $\al_5$, $\al_6$, $\al_7$, and $\al_{10}$ 
contribute only to the weak boson four-point couplings (QGC).

The typical size of the allowed range of 11 the 
chiral coefficients are hence:
\bea
O(\al_0,\al_1,\al_8) & \sim & 10^{-3} \,,\\
O(\al_2,\al_3,\al_9) & \sim & 10^{-1} \,,\\
O(\al_4,\al_5,\al_6,\al_7,\al_{10}) & \sim & 1 \,.\label{qgc1}
\eea
We will see in the following section, however, that the
QGC couplings in Eq. (\ref{qgc1}) are constrained severely
by the perturbation unitarity conditions for $\Lambda>1 \, \textrm{TeV}$.

\section{ Constraints on the chiral coefficients }
If we do not consider a particular class of the underlying
theory that leads to the electroweak symmetry breaking,
all the chiral coefficients 
are arbitrary parameters. In this pure phenomenological
viewpoint, we study constraints on their
magnitude from the electroweak precision data of the gauge boson
two-point functions, the TGC measurements, and from the perturbation
unitarity conditions from the weak boson scattering amplitudes.

In our analysis, the bounds on the chiral coefficients
listed in this section are determined at the scale $\mu=m_Z$.
Strictly speaking, the constraints on $\al_2$, $\al_3$, and $\al_9$
from the TGC measurements at LEP2 are obtained at the scale $\mu \approx 2 \, m_Z$, and those
on the $\al_4$, $\al_5$, $\al_6$, $\al_7$, and $\al_{10}$ are derived from the 
perturbative unitarity by assuming its validity   up to the scale  $\Lambda$.
However, since we consider the effects of these loosely constrained
chiral coefficients to the running of the most precisely
measured coefficients, $\al_0$, $\al_1$, and $\al_8$ only, the scale dependence
of all the other couplings give negligibly small effects on our results.
They can be considered as higher order corrections
of our leading order analysis.

Here we consider the most  general case by retaining all the operators
including those which  violate the custodial $SU(2)_c$ symmetry, since the
underlying dynamics can break it explicitly \cite{SSVZ}.  If
we impose the custodial symmetry, the following 
chiral coefficients vanishes:
\bea
\al_0, \al_6, \al_7,\al_8, \al_9, \al_{10} \equiv 0\,.
\label{cust}
\eea
We discuss the implication of the custodial symmetry in the latter
sections of this paper.

\subsection{Constraints from the two-point vertices}
By using three accurately measured quantities in  Table \ref{ggpv}, $1/\al_{\textrm{EM}}$,
$G_F$, and $m_Z$, we fix the vacuum expectation
value $v$, and the gauge couplings $g$ and ${g'}$.
The parameter $v$ is identified as
\bea
v(m_Z) = 1/\sqrt{\sqrt{2} G_F} = 246.26 \,\, \textrm{GeV}\,,
\eea
and the two gauge couplings are identified as
\begin{equation}
\begin{array}{rl}
g(m_Z) =& \,\,\,\,\,\,0.66\,,\\
g'(m_Z) =& \,\,\,\,\,\,0.36\,,
\end{array}
\label{eq:ggpv}
\end{equation}
by using the tree-level relations. We retain only two digits in
Eq. (\ref{eq:ggpv}), and do not consider their errors, because their small
variations do not affect our studies on the $S-T$ parameters in the
leading order.

To make a global fit with the oblique parameters $S$, $T$, and $U$ \cite{pandt},
we follow the strategy of Peskin and Wells \cite{Peskin:2001rw}, and
consider only three most precisely  measured quantities, the average value of 
charged leptonic partial decay width $\Gamma_\ell$ of $Z$,
the effective Weinberg mixing angle $\sin^2\theta^{\textrm{eff}}_{W}$,
and $m_W$. The latest results from the LEP, SLC, and Tevatron
\cite{unknown:2005em} are shown in Table \ref{STU}.

In the standard model, 
the relations among $\Gamma_\ell$,
$\sin^2\theta^{\textrm{eff}}_{W}$, $m_W$
and $S$, $T$, $U$ can be expressed as \cite{Hagiwara:1998}
\bea
m_W (\textrm{GeV})=& 80.377 \nnb \\
& - 0.288 \Delta S + 0.418 \Delta T + 0.337 \Delta U\,.\label{mw} \\
\Gamma_{\ell}(\textrm{GeV})=& 0.08395 \nnb \\
 & -0.00018 \Delta S + 0.00075 \Delta T\,, \label{gaell}\\
\sin^2\theta^{\textrm{eff}}_{\textrm{W}} =&0.23148  \nnb \\
 & +0.00359 \Delta S - 0.00241 \Delta T\,,\label{sin2w} 
\eea
where the central values are obtained 
by using the Zfitter 6.41 \cite{Arbuzov:2005ma} with $m_t^{ref}=175\textrm{GeV}$, 
$m_H^{ref}=100\,\textrm{GeV}$, $\al_s(m_Z)=0.1176$,
and $\delta \al_{5h}=0.0279$ as inputs.  The top and Higgs mass
dependence of the SM predictions are incorporated in the shifts
$\Delta S$, $\Delta T$, and $\Delta U$ \cite{Hagiwara:1998}.

In order to utilize the above formula designed for the theories with a Higgs boson to
theories like the EWCL without a Higgs boson, we follow the prescription of 
Bagger, Falk and Swartz, \cite{Bagger};
\beaa
\Delta S =& \Delta S_{\textrm{SM}}^{} - S_{\textrm{Higgs}}^{\textrm{ND}} + S\,\cma \\
\Delta T =& \Delta T_{\textrm{SM}}^{} - T_{\textrm{Higgs}}^{\textrm{ND}} + T\,\cma \\
\Delta U =& \Delta U_{\textrm{SM}}^{} - U_{\textrm{Higgs}}^{\textrm{ND}} + U\,.
\label{ewcl-fit}
\eeaa
Here  $S$, $T$, and $U$, are  the chiral couplings
in the EWCL, which do not have dependence on  $m_H$.
$S_{\textrm{Higgs}}^{\textrm{ND}}$, 
$T_{\textrm{Higgs}}^{\textrm{ND}}$, and $U_{\textrm{Higgs}}^{\textrm{ND}}$, 
are the Higgs boson contributions in the heavy Higgs limit and can be simply expressed as
\beaa
S_{\textrm{Higgs}}^{\textrm{ND}} =& - \,\,\,\,\,\,\,\,\,\,\,\,\,\frac{1}{6 \pi} \left[ \frac{5}{12} - \ln\left(\frac{m_H}{m_Z}\right)\right]\,,\\
T_{\textrm{Higgs}}^{\textrm{ND}} =&   \frac{1}{\cos^2\theta_W}  \frac{3}{8 \pi} \left[\frac{5}{12} - \ln\left(\frac{m_H}{m_Z}\right)\right]\,,\\
U_{\textrm{Higgs}}^{\textrm{ND}} =& 0\,.
\eeaa
By using the parameterization of $\Delta S_{\textrm{SM}}^{}$, $\Delta T_{\textrm{SM}}^{}$, 
and $\Delta U_{\textrm{SM}}^{}$ in Ref. \cite{Hagiwara:1998},
we observe that the $m_H$ dependence cancel accurately for $m_H>300 \textrm{GeV}$.
For definiteness, we set $m_H=500 \textrm{GeV}$ and find
\beaa
\Delta S_{\textrm{SM}}^{}- S_{\textrm{Higgs}}^{\textrm{ND}} =&\,\,\,\, 0.057 -0.007 x_t \,\cma \\
\Delta T_{\textrm{SM}}^{}- T_{\textrm{Higgs}}^{\textrm{ND}} =&-0.004 + 0.125 x_t + 0.003 x_t^2\,\cma \\
\Delta U_{\textrm{SM}}^{}- U_{\textrm{Higgs}}^{\textrm{ND}} =&-0.003 + 0.022 x_t \,,
\label{smstu}
\eeaa
where $x_t=(m_t - 175)/10$ parameterize the remaining $m_t$ dependence.

\begin{table}
\begin{ruledtabular}
\begin{tabular}{lcr}
inputs &  value     \\ \hline
$1/\al_{\textrm{EM}}^{}(m_Z)$ &   $127.87$ \\
$G_F$ & $1.166$ $\times$ $10^{-5} \textrm{GeV}^{-2}$\\
$m_Z$ & $91.18$ GeV \\
\end{tabular}
\caption{The inputs to fix $g(m_Z)$, $g'(m_Z)$, and $v(m_Z)$.}
\label{ggpv}
\end{ruledtabular}
\end{table}

It is now straightforward to find constraints of $S(m_Z)$, $T(m_Z)$,
and $U(m_Z)$ from the data of Table III, by using the parameterizations
Eqs. (\ref{mw}-\ref{smstu}), we find 
\begin{equation}
\begin{array}{rl}
S(m_Z) = &(-0.01\pm 0.10)\\
T(m_Z) = &(+0.09\pm 0.14)\\
U(m_Z) = &(+0.06\pm 0.13)
\end{array}
\rho =\begin{pmatrix}
1 &  & \cr 0.93 & 1 & \cr -0.55&-0.69&1
\end{pmatrix}\, ,
\label{stfit}
\end{equation}
where the $1 \sigma$ errors and their correlations
are given. In Fig. 1, we show the $1 \sigma$($39\%$ C.L.)
allowed region by a solid contour in the $S-T$ plane.

We can make one-to-one correspondences 
between the chiral coefficients 
$\al_1$, $\al_8$, $\al_0$ and the oblique parameters 
$S$, $T$, $U$ as in Ref. \cite{Appelquist:1993ka}
\begin{equation}
\begin{array}{rl}
\al_1(\mu) \equiv & - \frac{1}{16 \pi} \,\,\,\,S(\mu), \\
\al_0(\mu) \equiv &  \,\, \frac{\al_{\textrm{EM}}^{}}{2}\,\,\,\, T(\mu), \\
\al_8(\mu) \equiv & - \frac{1}{16 \pi} \,\,\,\,U(\mu)\,.
\end{array}
\label{stu2al108}
\end{equation}
Here the magnitudes of $\al_i(m_Z)$ are fixed by $S$, $T$, and $U$ at $\mu=m_Z$,
where $\al_{\textrm{EM}}^{}=1/137.36$ is the fine structure constant to
conform with the $T$ parameter definition of by Peskin and Takeuchi\cite{pandt}. The dependence
of the $\al_i(\mu)$ parameters on $\mu$ are determined by the RGE 
provided in section IV. Eq. (\ref{stu2al108}) can be
interpreted as the definition of the running $S$, $T$, and $U$ parameters
at $\mu\neq m_Z$. In terms of the chiral coefficients $\al_1(m_Z)$,
$\al_0(m_Z)$, and $\al_8(m_Z)$, the constraint Eq. (\ref{stfit}) gives
\begin{equation}
\begin{array}{rl}
\al_1(m_Z) = &(+0.02\pm 0.20) \times 10^{-2} \,,\\
\al_0(m_Z) = &(+0.03\pm 0.05) \times 10^{-2} \,,\\
\al_8(m_Z) = &(-0.12\pm 0.25) \times 10^{-2}\,,
\end{array}
\label{eq:al108}
\end{equation}
with the same correlation matrix.

\begin{figure}
\begin{center}
\includegraphics[height=8cm]{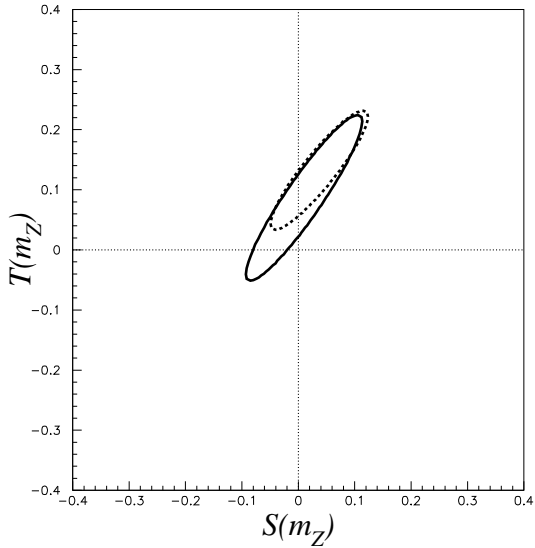}
\end{center}
\caption{The $S(m_Z)-T(m_Z)$ contours with $1 \sigma$ error from the electroweak data of Table
III. The solid-line contour shows the constraint without the custodial symmetry while the
dotted-line contour shows the constraint with the custodial symmetry ($U(m_Z)=0$).}
\label{fig1-j}
\end{figure}

When we impose the custodial symmetry. The constraint makes the
allowed area smaller, and the optimal values of
$S(m_Z)$ and $T(m_Z)$ moves up slightly, because of the positive
value of $U(m_Z)$ in the $3$-parameter fit and
the negative correlations in the third row of Eq. (\ref{stfit}).
We obtain
\begin{equation}
\begin{array}{rl}
S(m_Z) = &(+0.02\pm 0.09)\\
T(m_Z) = &(+0.13\pm 0.10)
\end{array}
\rho =\begin{pmatrix}
1 &  \cr 0.91 & 1
\end{pmatrix}\, .
\label{eq:st}
\end{equation}
The corresponding $1\sigma$ contour is shown in FIG. \ref{fig1-j} by a dotted-line contour.
In terms of the chiral coefficients, we impost $\al_8(m_Z) =0$ and find
\begin{equation}
\begin{array}{rl}
\al_1(m_Z) = &(-0.04\pm 0.17) \times 10^{-2} \,,\\
\al_0(m_Z) = &(+0.05\pm 0.04) \times 10^{-2} \,.
\end{array}
\label{eq:al108b}
\end{equation}
with the same correlation, $\rho=0.91$. We will use the constraints
Eq. (\ref{eq:st}) or equivalently Eq. (\ref{eq:al108b}) in the analysis when
the custodial $SU(2)_c$ symmetry is assumed.

These results roughly agree with those given in Ref. \cite{Bagger}, where the small
differences can be attributed to the changes in the input electroweak data.

\begin{table}
\begin{ruledtabular}
\begin{tabular}{lcr}
parameter & current value     \\ \hline
$m_W$  & $80.403\pm0.029$  GeV \\
$\Gamma_{\ell}$  & $83.984\pm0.086$ MeV \\ 
$\sin^2\theta_W^{\textrm{eff}}$ & $0.23152\pm0.00014$  \\
$m_t$ & $171.4 \pm 2.2 $ GeV \\
\end{tabular}
\caption{ Current values of the three best measured electroweak parameters
\cite{unknown:2005em} and $m_t$ \cite{topmass}
.}
\label{STU}
\end{ruledtabular}
\end{table}

\subsection{Constraints from the TGC 
\label{sub:tgc}}
There are three chiral coefficients $\al_2$, $\al_3$, $\al_9$ in the
EWCL which contribute to the triple gauge boson couplings (TGC) but not
to the two-point functions. 
The relations between the experimentally 
measured anomalous TGC  \cite{Hagiwara:1986vm} and the
three-point chiral coefficients
are given in Ref. \cite{Appelquist:1993ka}:
\begin{equation}
\begin{array}{rl}
\delta k_\gamma=&  \frac{e^2}{s^2} (-\al_1 - \al_8 + \al_2 + \al_3 + \al_9)  \,,\\
\delta k_z =&  \frac{1}{c^2-s^2} \al_0 + 
\frac{e^2}{c^2(c^2-s^2)} \al_1 +  \frac{e^2}{c^2} (\al_1 - \al_2)
\\ & + \frac{e^2}{s^2} (-\al_8 + \al_3 + \al_9)  \,,  \\
\delta g^1_Z =& \frac{1}{c^2-s^2} \al_0 + \frac{e^2}{c^2(c^2-s^2)} \al_1
 + \frac{e^2}{s^2 c^2} \al_3 \,.
\end{array}
\label{tri1}
\end{equation}
Because the constraints of $\al_1$, $\al_0$, and $\al_8$ in Eq. (\ref{eq:al108}) are much
more stringent than those of $\al_2$, $\al_3$, and $\al_9$ from 
the TGC measurements, we can simplify the above relations
to
\begin{equation}
\begin{array}{rl}
\delta k_\gamma =& (\al_2 + \al_3 + \al_9) g^2 \,, \\
\delta k_Z      =& - \al_2 {g^{\prime}}^2 + (\al_3 +  \al_9 ) g^2 \,,  \\
\delta g^Z_1    =&  \al_3 g_Z^2 \,,
\end{array}
\label{eq:tgc2al239}
\end{equation}
in our analysis. 

The unitarity bounds for anomalous TGC, $\delta k_\gamma$, $\delta k_z$ and $\delta g_1^Z$
derived \cite{Baur:1987mt} from 
processes $f^1 f^2 \rightarrow V^1 V^2$ are summarized as
\bea
|\delta k_\gamma| < \frac{1.86}{ \Lambda^2}\,,
|\delta k_Z | <  \frac{0.85}{ \Lambda^2}\,, 
|\delta g_1^Z | < \frac{0.87}{ \Lambda^2}\,. \label{tri-uni}
\eea
Here $\Lambda (\textrm{TeV})$ is the cut-off scale, up to which we request the validity of
the perturbation theory.
For $\Lambda = 1 \textrm{TeV}$, these unitarity bounds 
constrain the magnitudes of anomalous TGC  to be of order one. 
Hence, the unitarity constraints in Eq. (\ref{tri-uni}) can be neglected
when compared with the constraints from the TGC measurement 
even up to $\Lambda\sim 3 \textrm{TeV}$.

Constraints from the D0 experiment at the Tevatron 
$p {\bar p}$ collider \cite{Abazov:2005ys} on anomalous 
TGC  are not much stronger than the unitarity
bounds given in Eq. (\ref{tri-uni}) at $1$ TeV. It is the TGC measurements 
at LEP2 from the process $e^+ e^- \rightarrow W^+ W^-$ \cite{Heister:2001qt,Abbiendi:2003mk,
Achard:2004ji,Schael:2004tq}
that give the best constraints on these parameters.
We use the constraints on $\al_2$, $\al_3$, and $\al_9$ from the
LEP2 TGC measurements \cite{Heister:2001qt,Abbiendi:2003mk,
Achard:2004ji,Schael:2004tq} for the following three cases.

In {\bf case 1}, we impose the custodial symmetry on the anomalous dimensionless 
EWCL couplings, rendering $\al_9= 0$, {\it i.e.} Eq. (\ref{cust}). 
Eq. (\ref{eq:tgc2al239}) leads to a relation among  the  three TGC observables, 
\bea
\delta \kappa_Z = - \delta \kappa_\gamma \tan^2 \theta_w + \delta g^Z_1
\label{custodial}
\,,
\eea
In this scenario, we use the following results of the TGC measurements
by the LEP working group \cite{LEPEWWG},
each of which has been obtained from one-parameter fit:
\begin{equation}
\begin{array}{rl}
\delta  k_\gamma(m_Z) =& -0.03 \pm 0.05,\\
\delta  g^Z_1(m_Z)    =& -0.02 \pm 0.02,
\end{array}
\label{tgcdatacase1}
\end{equation}
Although there may be correlations between these two parameters in the two parameter
fit, it can be small \cite{LEPEWWG}. If we assume that the correlation between the errors is negligibly small,
then from Eq. (\ref{tgcdatacase1}) and $\al_9=0$, we find 
\begin{equation}
\begin{array}{rl}
\al_2(m_Z) =  \!\!\!&(-0.04\pm 0.12)\\
\al_3(m_Z) =  \!\!\!&(-0.03\pm 0.04)
\end{array}
\rho = \begin{pmatrix}
1 &  \cr -0.46 & 1 
\end{pmatrix}.
\label{eq:custal239}
\end{equation}

In {\bf case 2}, we adopt the two parameter-fitting result of the 
L3 collaboration \cite{Achard:2004ji}, 
which has also been obtained under the custodial symmetry (\ref{custodial}),
and reads:
\begin{equation}
\begin{array}{rl}
\de k_{\gamma}(m_Z) =  \!\!\!&(+0.16\pm 0.13)\\
\de g^Z_1(m_Z) =  \!\!\!&(-0.09\pm 0.05)
\end{array}
\rho = \begin{pmatrix}
1 & \cr -0.71 & 1
\end{pmatrix}.
\label{tgcdatacase2}
\end{equation}
In terms of the chiral coefficients, we find
\begin{equation}
\begin{array}{rl}
\al_2(m_Z) =  \!\!\!&(+0.54\pm 0.36)\\
\al_3(m_Z) =  \!\!\!&(-0.16\pm 0.10)
\end{array}
\rho = \begin{pmatrix}
1 &  \cr -0.82 & 1 
\end{pmatrix}.
\label{eq:l3al239}
\end{equation}
\begin{figure}
\begin{center}
\includegraphics[height=8cm]{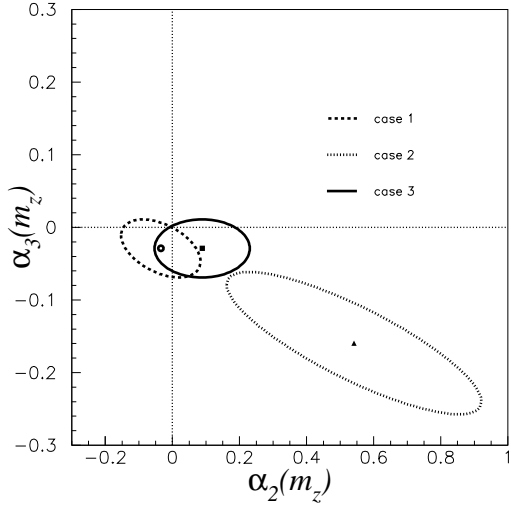}
\end{center}
\caption{The $\al_2-\al_3$ contours with $1 \sigma$ error ($39\%$ confidence level). 
In {\bf case 1}, the data is given in Eq. (\ref{tgcdatacase1}) and 
{\bf case 2}, the data is given in Eq. (\ref{tgcdatacase2}).
In both cases 1 and 2, the custodial symmetry condition $\al_9=0$ is imposed.
In {\bf case 3}, the data is given in Eq. (\ref{tgcdatacase3}) where 
$\al_9$ is taken as a free parameter in the fit.
}
\label{fig2-j}
\end{figure}

In the above two cases we set $\al_9 = 0 $ by appealing to
the custodial symmetry. In general, the custodial $SU(2)_c$ symmetry
that may explain the smallness of $\al_0$ does not necessarily
imply the suppression of $\al_9$. This motivates us to examine 
the third case where the chiral coefficients are analyzed in the
absence of the constraint from Eq.(\ref{custodial}).
 Since the experimental analysis for generic TGC without the constraint
cannot be found, we use 
the measurement of $k_Z(m_Z)$ from the L3 collaboration \cite{Achard:2004ji} 
as an input, even though the experimental analysis was carried out
by assuming the symmetry. 
Along with this analysis we choose the data on the other two  TGC observables from the 
the LEP combined limits, Eq. (\ref{tgcdatacase1}), 
on $\delta  k_\gamma(m_Z)$ and $\delta  g_1^Z(m_Z)$ \cite{LEPEWWG},
as the {\bf case 3}:
\begin{equation}
\begin{array}{rl}
\delta  k_Z(m_Z)      =& -0.08 \pm 0.06\,,\\
\delta  k_\gamma(m_Z) =& -0.03 \pm 0.05,\\
\delta  g^Z_1(m_Z)    =& -0.02 \pm 0.02.
\end{array}
\label{tgcdatacase3}
\end{equation}
Although the above limits should be much stronger than the true constraints when
all the three TGC are allowed to vary freely, we adopt the above bounds as inputs
of our general analysis without the custodial symmetry constraints. 
Again by neglecting the correlations
among the errors of Eq. (\ref{tgcdatacase3}), we find 
\begin{equation}
\begin{array}{rl}
\al_2(m_Z) =  \!\!\!&(+0.09\pm 0.14)\\
\al_3(m_Z) =  \!\!\!&(-0.03\pm 0.04)\\
\al_9(m_Z) =  \!\!\!&(-0.12\pm 0.12)
\end{array}
\rho = \begin{pmatrix}
1 &  & \cr 0.00 & 1 & \cr -0.65&-0.32&1
\end{pmatrix}.
\label{eq:al239}
\end{equation}
The bound is too stringent 
because the individual results of Eq. (\ref{tgcdatacase3}) are obtained under
the constraint of Eq. (\ref{custodial}) and also because the possible
correlations among those measurements are neglected. However,
it serves our purpose of showing the possible
impacts of custodial $SU(2)_c$ symmetry violation.

In all the above three cases, we observe
that $\al_3$ is more stringently constrained 
by the LEP2 data than $\al_2$ and $\al_9$.
FIG. \ref{fig2-j} shows the $1 \sigma$ error contour for $\al_2$-$\al_3$ for 
these three 
cases. We note that the central values of 
the L3 data deviate significantly from the prediction of the standard model.

\subsection{Perturbative unitarity constraints}
The most stringent theoretical constraint for the four-point chiral
coefficients comes from the partial wave unitarity conditions of the 
longitudinally polarized vector boson
scattering amplitudes. These scattering amplitudes grow with energy if 
there are no new resonances up to the cut off scale $\Lambda$  \cite{Cornwall:1973tb}.

Here we only consider the processes with 
$J=0$ channels and find 
\begin{eqnarray}
&&\hskip -0.5 cm 
\begin{array}{rl}
|4 \al_4 + 2\al_5 |  < & {3 \pi} \frac{v^4 }{ \Lambda^4},
\\
| 3 \al_4 + 4 \al_5 |  < &{3 \pi } \frac{v^4 }{ \Lambda^4},
\\
|\al_4 + \al_6 + 3 (\al_5 + \al_7) | < & {3 \pi } \frac{v^4 }{ \Lambda^4},
\\
|2 (\al_4 +\al_6) + \al_5  +\al_7 |  <  & {3 \pi } \frac{v^4 }{ \Lambda^4},
\\
| \al_4 + \al_5 + 2 (\al_6 + \al_7 + \al_{10}) |  < &
\frac{6 \pi }{5 } \frac{v^4 }{ \Lambda^4}.
\end{array}
\label{eq:al456710}
\end{eqnarray}
The five constraints are obtained from $W^+_L W^+_L \rightarrow W^+_L W^+_L$,
$W^+_L W^-_L \rightarrow W^+_L W^-_L$, $W^+_L W^-_L \rightarrow Z_L Z_L$,
$W^+_L Z_L   \rightarrow W^+_L Z_L $, and $Z_L Z_L \rightarrow Z_L Z_L$, 
respectively.
In our analysis,the  effects of the terms proportional to  $v^2/\Lambda^2$ 
are found to be negligibly small and hence are dropped.

\begin{table}
\begin{ruledtabular}
\begin{tabular}{| c | c | c | }
 Chiral Coefficients & Central values & Error bars  ($\pm$)  \\ \hline
$\al_0(m_Z)$  &   $+0.0003 $       &     $ 0.0005 $    \\
$\al_1(m_Z)$  &   $+0.0002 $       &     $ 0.0020 $    \\
$\al_2(m_Z)$  &   $-0.09$                      &     $ 0.14$ \\
$\al_3(m_Z)$  &   $+0.03$                      &     $ 0.04$ \\
$\al_4(m_Z)$  &                                &     $ \sim  \pi \frac{v^4}{\Lambda^4}$ \\
$\al_5(m_Z)$  &                                &     $ \sim  \pi \frac{v^4}{\Lambda^4}$ \\
$\al_6(m_Z)$  &                                &     $ \sim  \pi \frac{v^4}{\Lambda^4}$ \\
$\al_7(m_Z)$  &                                &     $ \sim  \pi \frac{v^4}{\Lambda^4}$ \\
$\al_8(m_Z)$  &   $-0.0012 $       &     $ 0.0025 $    \\
$\al_9(m_Z)$  &   $+0.13$                      &     $ 0.12$    \\
$\al_{10}(m_Z)$  &                             &     $ \sim  \pi \frac{v^4}{\Lambda^4}$    \\
\end{tabular}
\end{ruledtabular}
\caption{ Summary on our knowledge of 
chiral coefficients, from Eq. (\ref{eq:ggpv}), Eq. (\ref{eq:al108}),
Eq. (\ref{eq:al239}) (here we take the most stringent bounds), and Eq. (\ref{eq:al456710}). }
\label{data}
\end{table}

Recently, Distler {\it et. al.} \cite{Distler:2006if} 
found that from the amplitudes of the scattering processes $Z_L Z_L \rightarrow Z_L Z_L$ 
and $W_L Z_L \rightarrow W_L Z_L$,
by using dispersion relations and by assuming Lorentz invariance, analyticity, unitarity,
and custodinal symmetry,
it is possible to derive the lower bounds for the 
chiral parameters $\al_4$ and $\al_5$, which read
\begin{equation}
\begin{array}{rl}
\al_5 + 2 \al_4 & \geq \frac{1}{96 \pi^2} \times 1.08  \,, \\
\al_4 & \geq \frac{1}{96 \pi^2} \times 0.31 \,.
\end{array}
\label{lowbound}
\end{equation}
To quote these number, we set $\mu=v$ and $\delta =1/5$.

We now summarize the current 
bounds on the chiral coefficients of the EWCL in Table \ref{data}. Among
the six terms that contribute to the two-point functions,
the first three coefficients, $g$, $g'$, and $v$, 
are fixed by $\al_{\textrm{EM}}^{}(m_Z)$, $G_F$, and $m_Z$ 
in Table II, and
the remaining three 
chiral coefficients, $\al_0$, $\al_1$, and $\al_8$ are
constrained from the Z pole data, the precise W mass measurement, and the top quark
mass measurement, see Table III.
The three three-point chiral coefficients, $\al_2$, $\al_3$, and
$\al_9$ are determined from the LEP2 W pair production measurements, and the 
five four-point chiral coefficients, $\al_4$, $\al_5$, $\al_6$, 
$\al_7$, and $\al_{10}$ are constrained by the five perturbative unitarity conditions
of Eq. (\ref{eq:al456710})
if there is no new resonances up to the scale $\Lambda$.
 
We
observe from Table IV that the two-point coefficients $\al_0$, $\al_1$, $\al_8$ are constrained
to be $\sim 10^{-3}$, the TGC coefficients $\al_2$, $\al_3$, and $\al_9$
can be $\sim 10^{-1}$, while the remaining coefficients should be smaller than $\pi (v/\Lambda)^4$,
or $10^{-2}$ for $\Lambda \sim 1 \textrm{TeV}$.

It may be instructive to compare the above constraints with the corresponding
ones of the QCD chiral theory.
Reference \cite{Gasser:1984gg} presents the 
fit of these dimensionless couplings in the chiral perturbation theory
describing low energy QCD. 
The constraints found in reference \cite{Harada:2003jx} 
are $L_{9}(m_\eta)=(7.4\pm0.7 \times 10^{-3})$ and $L_{10}(m_\eta)=(-6.0\pm0.7 \times 10^{-3})$, 
which correspond to  $\al_2$($=\al_3$) and  $\al_1$ respectively, in 
the electroweak theory. 

The large and positive value of $\al_1$ at $\Lambda\sim 1 \textrm{TeV}$, favored by
the electroweak data, has been confronted against the large and negative
value of $L_{10}(m_{\eta})$, which led to
the assertion \cite{pandt} that the electroweak symmetry breaking does not mimic QCD.
Although strictly QCD-like theories may also give $\al_2 \sim \al_3 \sim 10^{-2}$ as $L_9(m_\eta)$,
we examine possible implications of models with $\al_2, \al_3, \al_9 \sim 10^{-1}$,
which are still allowed by the present TGC measurements.

\section{ The $\beta$ function of chiral coefficients at one-loop level}

In order to find the connection between $S(m_Z)-T(m_Z)$ and
$S(\Lambda)-T(\Lambda)$, by using the method developed in \cite{dhy1}, 
we extend the previous one-loop RGE \cite{Herrero:1993nc} of the EWCL 
by including the three-point and four-point chiral coefficients 
in the $\beta$ functions of
$S$ and $T$ (equivalently, the $\beta$ functions of the 
chiral coefficients $\al_1$ and $\al_0$).
 In the Wilsonian renormalization group concept, at the one-loop level, 
all dimensionless chiral coefficients $\al_i$ should
run in a logarithmic way from $m_Z$ to $\Lambda$, just like the gauge couplings,
due to the screening effects of the active quantum degree of freedoms (such as
the Goldstone particles, the gauge bosons, and the ghosts). This logarithmic
running is obtained by using the dimensional regularization in our
calculation.

The RGE for the gauge couplings $g$ and $g'$, for the vev term $v^2$, as well as
for the chiral coefficients $\al_i $ can be simply expressed as
\begin{eqnarray}
&&\hskip -0.5 cm 
\begin{array}{rl}
\frac{d}{dt} g = & \frac{1}{8 \pi^2} \beta_{g} \,, \\
\frac{d}{dt} g^\prime = & \frac{1}{8 \pi^2} \beta_{g^\prime} \,,\\
\frac{d}{dt} v^2 =  &\frac{1}{8 \pi^2} \beta_{v^2} \,,\\
\frac{d}{dt} \al_i =  &\frac{1}{8 \pi^2} \beta_{\al_i} \,,
\end{array}
\label{eq:beta}
\end{eqnarray}

Where $t=\ln (\mu/m_Z)$. The $\beta_{g,g',v^2,\al_i}$ are 
the beta functions for the running 
of $g$ and $g'$, $v^2$, as well as $\al_i $. Below we list the $\beta$ functions 
of the chiral coefficients of the EWCL at one-loop.
Technical details of our calculation is described in Appendix A.
The gauge fixing terms and
the treatment of the ghost terms are presented in Appendix B.

We organize and group  these $\beta$ functions
in the order of their contributions to the multi point functions.
The $\beta$ functions are written for those operators which contribute to the
two point functions ($\al_0$, $\al_1$, and $\al_8$), 
three point functions ($\al_2$, $\al_3$, and $\al_9$),
and four point functions ($\al_4$, $\al_5$, $\al_6$, $\al_7$, and $\al_{10}$). 
In the standard derivative power counting rule, these
terms are counted as $O(p^6)$ order effects.

We first write $\beta_{g,\,g^\prime}$  for the gauge couplings.
\bea
\label{rge0}
\beta_{g}&=&
\frac{g^3}{2} \mbbkf{ -\nkf{ \frac{29}{4} } - \frac{\al_0 }{ 6} - 
    \my{\al_1} {\my{g'}}^2 - 
    4  \my{\al_8} g^2  \wsep  + 
    \frac{5 \my{\al_2} {\my{g'}}^2}{6} - 
    \my{\al_3} \sbkf{ \frac{28 g^2}{3} + 
       \frac{{\my{g'}}^2}{2} } - 
    \frac{13 \my{\al_9} g^2}{6}   } 
\,,
\label{rgeg}
\eea
\bea
\beta_{g'}&=&
\frac{{\my{g'}}^3 }{2}\sbbkf{ \frac{1}{12} - \frac{\al_0 }{ 3} - 2 \my{\al_1} g^2  
      \wsep - \my{\al_2} g^2 + \frac{5 \my{\al_3} g^2}{3} }  
\,.
\label{rgegp}
\eea
It is noted  that the $\beta_g$ and $\beta_{g^\prime} $ both have constant terms which 
are expected by naively counting of the active bosonic degrees of freedom. 
Compared to the SM $\beta$ functions, only the physical Higgs boson contributions
are absent. We have not 
included the contribution of active fermionic degree of freedoms
which can be fixed in a straightforward manner.
Also we find that the three-point chiral coefficients $\al_2$, $\al_3$, and $\al_9$,
contribute to the $\beta$ functions of the gauge couplings, and
can modify the asymptotic behavior of gauge couplings,
which should match the gauge couplings of the ultra-violet
theory at the scale $\Lambda$.

The $\beta$ function of $v^2$ is given as
\bea
\beta_{v^2}&=&
\frac{3 g^2}{2} + \frac{3 {\my{g'}}^2}{4}  + \al_0 (\frac{3}{2}  g^2 -3 {g'}^2 )  \wsep - 
    \frac{3\my{\al_1} g^2 {\my{g'}}^2}{2}  + 
   \frac{3 \my{\al_8} g^4}{4}  
\wsep + \my{\al_2} \sbbkf{ 3 g^2 {\my{g'}}^2 - 
     \frac{3 {\my{g'}}^4}{2} }  - 
  \my{\al_3} \sbbkf{ 2 g^4 - 3 g^2 {\my{g'}}^2 } 
 \wsep - \my{\al_9} \sbbkf{ g^4 - 
     \frac{3 g^2 {\my{g'}}^2}{2} } \wsep  -
  \my{\al_4} \sbbkf{ \frac{21}{4} g^4   +
     \frac{3 g_Z^4}{4} } - 
  \my{\al_5} \sbbkf{ \frac{15}{2}  g^4 +
     3 g_Z^4 } \wsep -
   \frac{3 \my{\al_6}  g_Z^4 }{4}  - 
   3 \my{\al_7}  g_Z^4     
\,.
\label{rgev2}
\eea
For $\beta_{v^2}$ we  notice that  it has no dependence on $\al_{10}$.
This can be attributed to the fact that the use of dimensional regularization 
results in retaining only the 
logarithmic running of $v^2$.

The  $\beta$ functions of chiral couplings, $\al_0$, $\al_1$, and $\al_8$,
are given as
\bea
\beta_{\al_0}&=&
 \frac{3 {\my{g'}}^2}{8}  + \al_0 ( \frac{9  g^2 }{ 4} - \frac{9 {\my{g'}}^2}{ 4 } )
 \wsep -   \frac{3  \my{\al_1} g^2 {\my{g'}}^2}{4} 
  +  \frac{3 \my{\al_8} g^4}{8}  \wsep + 
  \my{\al_2} \sbbkf{ \frac{3 g^2 {\my{g'}}^2}{2} - 
     \frac{3 {\my{g'}}^4}{4} } + 
   \frac{3  \my{\al_3} g^2 {\my{g'}}^2}{2} \wsep + 
  \my{\al_9} \sbbkf{ -\frac{g^4}{2} + 
     \frac{3 g^2 {\my{g'}}^2}{4} }  \wsep +
  \my{\al_4} \sbbkf{ \frac{15 g^2 {\my{g'}}^2}{4} + 
     \frac{15 {\my{g'}}^4}{8} }  + 
  \my{\al_5} \sbbkf{ \frac{3 g^2 {\my{g'}}^2}{2} +
     \frac{3 {\my{g'}}^4}{4} } \wsep +
  \my{\al_6} \sbbkf{ \frac{3 g^4}{4} + 
     \frac{33 g_Z^4}{8} }  +
  \my{\al_7} \sbbkf{ 3 g^4 + 3 g_Z^4 }  \wsep   + 
    \frac{9 \my{\al_{10}} g_Z^4}{2} 
\,.
\label{rgeal0}
\eea
\bea
\beta_{\al_1}&=&
\frac{1}{12} + 4 \my{\al_1} g^2 - \my{\al_8} g^2 
\wsep  - \frac{5 \my{\al_2} g^2}{2} + 
  \frac{5 \my{\al_3} g^2}{6} - 
  \frac{\my{\al_9} g^2}{2}\,,
\label{rgeal1}
\eea
\bea
\beta_{\al_8}&=& \frac{\al_0 }{ 2} + 
  \my{\al_1} {\my{g'}}^2 +
 12 \my{\al_8} g^2 \wsep  - 
  \frac{5 \my{\al_2} {\my{g'}}^2}{6} + 
  \frac{\my{\al_3} {\my{g'}}^2}{2}
- \frac{17 \my{\al_9} g^2}{6} \,,
\label{rgeal8}
\eea
We remark here that the three-point chiral coefficients $\al_2$, $\al_3$, and $\al_9$,
affect the running behavior of the two-point chiral coefficients $\al_1$, $\al_8$, and $\al_0$, 
parameters. We observe that the $\beta$ functions for $\al_1$ and $\al_8$ do not 
contain the  
four-point chiral coefficients $\al_4$, $\al_5$, $\al_6$, $\al_7$, 
and $\al_{10}$ but $\al_0$ does.
Interpreting this exception in  term of Feynman diagrams, we observe that $\al_1$ and $\al_8$
only receives the  radiative corrections through the diagram of FIG. (3a) while $\al_0$
receives the  radiative corrections from both FIG. (3a) and FIG. (3b).
It is the diagram FIG. (3b) which renders the entry of 
the four-point chiral coefficients in  the $\beta$ function
of $\al_0$.

\begin{figure*}
\begin{minipage}[t]{6.0cm}
     \includegraphics[height=3.5cm]{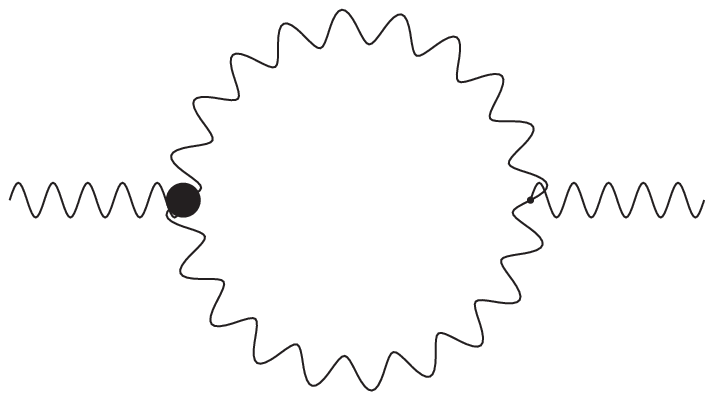}
     \mbox{ }\hfill\hspace{1.0cm}(3a)\hfill\mbox{ }
     \end{minipage}
     \hspace{2cm}
     \begin{minipage}[t]{6.0cm}
     \includegraphics[height=3.5cm]{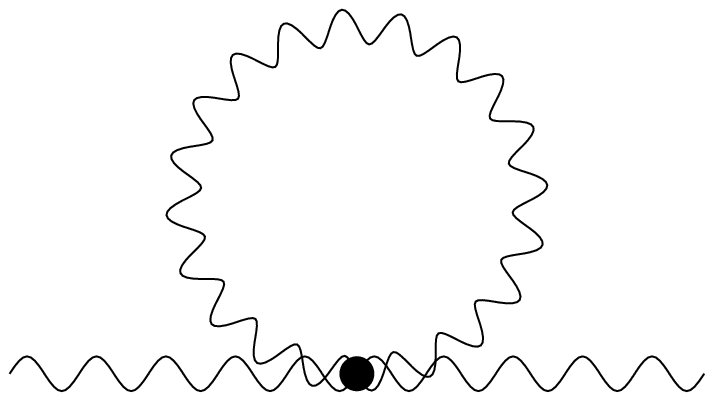}
     \mbox{ }\hfill\hspace{1.0cm}(3b)\hfill\mbox{ }
     \end{minipage}
     \caption{ Feynman diagrams for two-point functions. 
Wavy line loops should include vector, Goldstone, and ghost loops.
The solid circles show the chiral operators ${\cal L}_i$.
}
\label{fig3-j}
\end{figure*}

The $\beta$ functions of the three-point chiral coefficients $\al_2$, $\al_3$
and $\al_9$, are given as
\bea
\beta_{\al_2}&=&
\frac{1}{24} + \frac{\al_0}{6} - \frac{ \my{\al_1} g^2 }{2}
\wsep + \my{\al_2} \sbbkf{2 g^2 + \frac{{ g'}^2}{12}}  
  + \frac{5 \my{\al_3} g^2}{12} - 
  \frac{\my{\al_9} g^2}{12} 
\wsep + \al_4 \sbbkf{\frac{g^2}{4} + \frac{{g'}^2}{2}} - \al_5 \sbbkf{\frac{g^2}{2} 
- {g'}^2} 
\wsep - \frac{\al_6 {g'}^2}{2} - \al_7 {g'}^2\,,
\label{rgeal2}
\eea
\bea
\beta_{\al_3}&=&
\frac{1}{24} - \frac{\al_0}{6} - \frac{ \my{\al_1} {g'}^2 }{4}
\wsep + \frac{\my{\al_2} {g'}^2 }{6}
  + \my{\al_3} \sbbkf{\frac{59  g^2}{12}  + \frac{{g'}^2}{4}}- 
  \frac{5\my{\al_9} g^2}{3} 
\wsep + \al_4 \sbbkf{\frac{5 g^2}{4} + \frac{3 {g'}^2}{8}} - \al_5 \sbbkf{\frac{5 g^2}{2} 
- \frac{{g'}^2}{4}} 
\wsep + \al_6\sbbkf{ \frac{5 g^2}{4} + \frac{3 {g'}^2}{8} } - \al_7 \sbbkf{\frac{5 g^2}{2} 
- \frac{{g'}^2}{4}}\,,
\label{rgeal3}
\eea
\bea
\beta_{\al_9}&=&
 \frac{\al_0}{2} - \frac{ \my{\al_1} {g'}^2 }{4} - \frac{ 3 \my{\al_8} g^2 }{4} 
\wsep + \frac{\my{\al_2} {g'}^2 }{12}
  + \frac{ 5 \my{\al_3}  {g'}^2}{4} + 
  \frac{119 \my{\al_9} g^2}{12} 
\wsep - \frac{7 \al_4 {g'}^2}{8}   
- \frac{5 \al_5 {g'}^2}{4}
\wsep - \al_6\sbbkf{ \frac{5 g^2}{4} + \frac{7 {g'}^2}{8} } 
- \al_7 \sbbkf{- \frac{5 g^2}{2} 
+ \frac{5 {g'}^2}{4}}\,,
\label{rgeal9}
\eea
The $\beta$ functions of the four-point chiral coefficients, $\al_4$,
$\al_5$, $\al_6$, $\al_7$, and $\al_{10}$ are
given as
\bea
\beta_{\al_4}&=&
 -\frac{1}{12} - \al_0 - \frac{5 \my{\al_2} {g'}^2 }{6}
  - \frac{  \my{\al_3}  g^2}{2} -
  \frac{ \my{\al_9} g^2}{6} 
\wsep + \al_4 \sbbkf{ \frac{11 g^2}{2} + 6 {g'}^2 }     
+\al_5\sbbkf{ 2 g^2 + 3  {g'}^2}
\wsep + \al_6\sbbkf{- \frac{g^2}{2} + \frac{7 {g'}^2}{2} } 
+ 2 \al_7 {g'}^2 \,,
\label{rgeal4}
\eea
\bea
\beta_{\al_5}&=&
 -\frac{1}{24} + \frac{\al_0}{2} 
 - \frac{ \my{\al_2} {g'}^2 }{3}
 - \frac{  \my{\al_3}  g^2}{2} - 
  \frac{ \my{\al_9} g^2}{3} 
\wsep - \al_4 \sbbkf{ g^2 + \frac{15 {g'}^2}{4}  }    
+\al_5\sbbkf{ g^2 - \frac{3  {g'}^2}{2} }
\wsep - \al_6\sbbkf{ \frac{g^2}{4} + \frac{3 {g'}^2}{4} } 
+ \al_7 \sbbkf{ - \frac{3 g^2}{2} + \frac{3 {g'}^2}{2} }\,,
\label{rgeal5}
\eea
\bea
\beta_{\al_6}&=&
 \frac{3 \al_0}{2}  - \my{\al_2} {g'}^2
 +\frac{25 \al_4  {g'}^2}{4}  
+ \frac{7 \al_5 {g'}^2 }{2}
\wsep - \al_6\sbbkf{ \frac{25 g^2}{4} - \frac{5 {g'}^2}{4} } 
+ 2 \al_7 g^2
\wsep + \al_{10} \sbbkf{-g^2 + 5 {g'}^2} \,,
\label{rgeal6}
\eea

\bea
\beta_{\al_7}&=&
- \frac{3 \al_0}{4}  - \frac{3 \my{\al_2} {g'}^2}{4}
+ \frac{ \my{\al_9} {g'}^2}{4}
\wsep +\frac{41 \al_4  {g'}^2}{8}  +   
+ \frac{11 \al_5 {g'}^2 }{4}
\wsep + \al_6\sbbkf{ - \frac{5 g^2}{8} + \frac{33 {g'}^2}{8} } 
+ \al_7 \sbbkf{  \frac{13 g^2}{4} + \frac{7 {g'}^2}{4} }
\wsep 
+ \al_{10} \sbbkf{-2 g^2 + 4 {g'}^2} \,,
\label{rgeal7}
\eea

\bea
\beta_{\al_{10}}&=&
 \frac{ 17 \al_6 {g'}^2}{4} 
+ \frac{ 7 \al_7 {g'}^2}{2} 
+ 9 \al_{10} \sbbkf{g^2 - {g'}^2} \,.
\label{rgeal10}
\eea

In calculating the above $\beta$ functions, 
we have performed the following consistency checks: 
{\bf (1)} We ensure that when all the anomalous couplings are set 
to vanish, our calculation
reduces   not only to the standard  $\beta$ functions for gauge couplings 
but also to the constant terms in the  $\beta$ functions for the
chiral coefficients \cite{Herrero:1993nc}.
{\bf (2)} The electromagnetic symmetry $U_{\textrm{EM}}^{}(1)$ is checked and 
found to hold in our calculation at each step.
{\bf (3)} Whenever ${g'}$ is set to vanish, $Z$ can be expressed in terms of  $W^3$, 
and  similarly $W^{\pm}$ can be
expressed in terms of $W^1$ and $W^2$. Thus we find that the 
global $SU(2)_c$ custodial symmetry explicitly validated 
in our calculation at each step.
{\bf (4)} When we only keep $\al_0$ in the $\beta$ functions, our
results agree with those obtained by M. Tanabashi \cite{private}
and in Ref. \cite{Chivukula:2007ic}. {\bf (5)} Using the alternative
parameterization for Goldstone boson as
\bea
U=\exp\bbra{ {\bf i} \frac{ 2 \, \xi_3 \, T^3}{v} } \exp\bbra{{\bf i}  \frac{ 2 \xi^1 T^1 +  2 \xi^2 T^2}{ v} }\,,
\eea
our computational method yields the same answer.

\section{Current experimental uncertainty in $S(\Lambda)-T(\Lambda)$}
In this  section  we analyze the current experimental uncertainty of $S(\Lambda)-T(\Lambda)$
obtained from the constraints at the $m_Z$ scale listed in Table \ref{data}. First, we 
analyze the experimental uncertainty of  $S(\Lambda)-T(\Lambda)$ induced
by the uncertainty of the three-point chiral coefficients.
Second, we will analyze the experimental uncertainty
of $T(\Lambda)$ induced by both the uncertainty of 
the three-point and the four-point chiral coefficients.

Numerically, we find that
$S(\Lambda)$, $T(\Lambda)$, and $U(\Lambda)$, ( or $\al_1(\Lambda)$, $\al_0(\Lambda)$, 
and $\al_8(\Lambda)$ via Eq. (\ref{stu2al108}) ) are not
sensitive to the running of the other chiral coefficients and
that of the gauge couplings. Therefore we can use linear 
approximation of the RGE solutions:
\begin{equation}
\begin{array}{rl}
S(\Lambda) =& S(m_Z) - \frac{2}{\pi} \beta_{\al_1} \ln \frac{\Lambda}{m_Z}\,, \\
T(\Lambda) =& T(m_Z) + \frac{1}{4 \pi^2 \al_{\textrm{EM}^{}}} \beta_{\al_0} \ln \frac{\Lambda}{m_Z}\,, \\
U(\Lambda) =& U(m_Z) - \frac{2}{\pi} \beta_{\al_8} \ln \frac{\Lambda}{m_Z}\,.
\end{array}
\label{rgesol}
\end{equation}

Here we explain the basic difference
between the experimental uncertainty analysis of $S-T$ 
given in \cite{pdg2006} and ours.
The $S-T$ uncertainty contour figures given in \cite{pdg2006}
are obtained in the minimal standard
model with a Higgs boson as a regulator, along with the  introduction of
three extra two point operators, ${\bar {\cal L}}_0$, 
${\bar {\cal L}}_1$, and ${\bar {\cal L}}_8$, to
describe the new physics effects. 
In this analysis, although the central values of $S$, $T$, and $U$
vary with the Higgs boson mass, their error bars are insensitive to $m_H$ and
are determined by the experimental errors.
To have an analogy with the prediction of the QCD-like theories, 
the experimental values of $S-T$ are determined by choosing $m_H = 1$ TeV ($1 \textrm{TeV}$
is assumed to be the compositeness scale).

In our uncertainty analysis, we do not use Higgs as regulator
but adopt the dimensional
regularization and the ${\overline {\rm MS}}$ scheme.
The uncertainties of $S(m_Z)$, $T(m_Z)$, $U(m_Z)$ and their
correlations are then determined by the electroweak data listed
in the Table \ref{data}. Therefore, the errors at the scale $\mu=m_Z$
are essentially the same as those of Ref. \cite{pdg2006}. In our analysis, however,
the error bars of $S(\Lambda)-T(\Lambda)$ become
larger and larger when we extrapolate the low energy constraints to high energy
scale, since more and more uncertainty from the other operators creeps in the 
computation. 

\subsection{The experimental uncertainty of $S(\Lambda)-T(\Lambda)$ from  
the anomalous TGC  measurement}

The parameters, $S(\Lambda)$, $T(\Lambda)$ and $U(\Lambda)$ are the 
values of parameters $S$, $T$ and $U$ at the matching scale $\Lambda$,
where the EWCL matches with the underlying 
fundamental theories such as the QCD-like models, 
extra dimension models, Higgsless models, etc.
In the perturbation method, $S(\Lambda)$, $T(\Lambda)$, $U(\Lambda)$ 
and $S(m_Z)$, $T(m_Z)$, $U(m_Z)$ are connected
by the improved renormalization group equations
Eq.(\ref{rgeal1}), Eq.(\ref{rgeal0}), and Eq.(\ref{rgeal8}), respectively.

\begin{figure}
\begin{center}
\includegraphics[height=8cm]{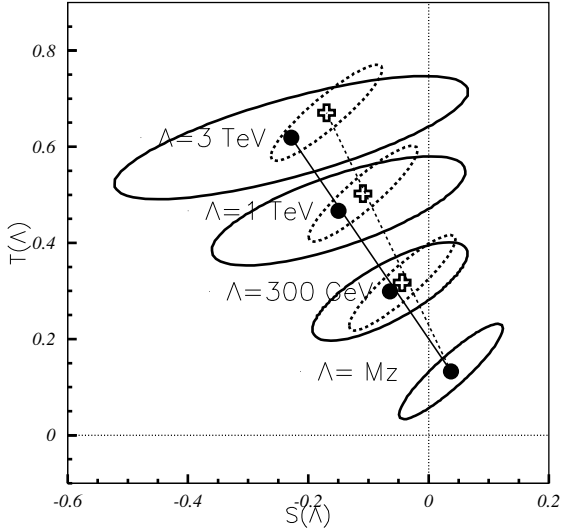}
\end{center}
\caption{ The $S(\Lambda)-T(\Lambda)$ contours for various cutoff scales $\Lambda$.
The solid contours are obtained when the TGC data is taken from the
{\bf case 1}, Eq. (\ref{tgcdatacase1}), and $U(m_Z)=+0.00$. 
The dashed contours are obtained when all the three and four point chiral
coefficients are set to be zero.} 
\label{fig4-j}
\end{figure}

\begin{figure}
\begin{center}
\includegraphics[height=8cm]{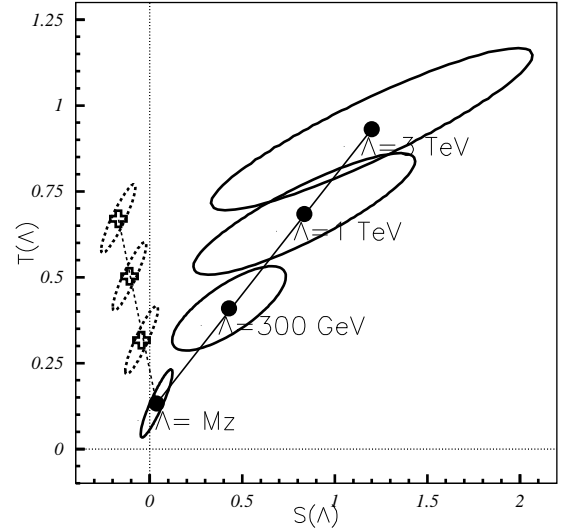}
\end{center}
\caption{ The $S(\Lambda)-T(\Lambda)$ contours for various cutoff scales $\Lambda$.
The solid contours are obtained when the TGC data is taken from the
{\bf case 2}, Eq. (\ref{tgcdatacase2}), and $U(m_Z)=+0.00$. 
The dashed contours are obtained when all the three and four point chiral
coefficients are set to be zero.} 
\label{fig5-j}
\end{figure}

\begin{figure}
\begin{center}
\includegraphics[height=8cm]{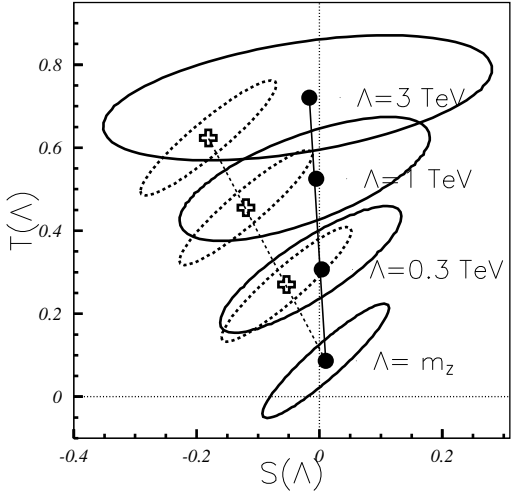}
\end{center}
\caption{ The $S(\Lambda)-T(\Lambda)$ contours for various cutoff scales $\Lambda$ when
the custodial symmetry is not imposed on the chiral coefficients.
The solid contours are obtained when the TGC data is taken from the
{\bf case 1}, Eq. (\ref{tgcdatacase3}) and $S$, $T$, and $U$ at $\mu=m_Z$
are constrained by Eq. (\ref{stfit}). 
The dashed contours are obtained when all the three and four point chiral
coefficients are set to be zero.
} 
\label{fig6-j}
\end{figure}

With the $S$, $T$, $U$ values at $\mu=m_Z$ in subsection (IIIA)
and the three-point chiral coefficients at $\mu=m_Z$ for all the three cases as given
in subsection (\ref{sub:tgc}), we evolute the 
values of $S$, $T$, $U$ up to the cutoff scale $\Lambda$.
The results are shown in Fig. (\ref{fig4-j}) for 
the {\bf  case 1} with the constraints in Eq. (\ref{eq:custal239}),
Fig. (\ref{fig5-j}) for the {\bf case 2} 
with the constraints in Eq. (\ref{eq:l3al239}), 
and Fig. (\ref{fig6-j}) for the {\bf case 3} 
with the constraints in Eq. (\ref{eq:al239}). 
In Figs. (\ref{fig4-j}) and (\ref{fig5-j}) for the {\bf case 1} and {\bf case 2},
we set all the coefficients that violate the custodial symmetry
to zero at the scale $m_Z$ except $\al_0$, and 
therefore we use Eq. (\ref{eq:st}) for $U(m_Z)=0$ as the input.
Because the hypercharge gauge interactions violate the
custodial symmetry, we cannot impose the condistions at all the scales.
According to Eq. (\ref{rgeal8}), however, 
$U(\Lambda)$ remains negligibly small at $\Lambda=1\, TeV$,
$U(1\,\textrm{TeV}) = -0.05 \pm 0.20$ for the {\bf case 1} and 
$U(1\,\textrm{TeV}) = -0.08 \pm 0.55$ for the {\bf case 2}.
In Fig. (\ref{fig6-j}) for the {\bf case 3}, we allow all the coefficients to vary,
and use Eq. (\ref{stfit}) as the input.
Results for $\Lambda$ at $300$ GeV, $1$ TeV, and $3$ TeV are shown in these figures.

The dashed line contours correspond to the analysis without  
the contributions of the three-point chiral coefficients,
{\it i.e.} they include only the 1 $ \sigma$ error of the two-point chiral 
coefficients. Therefore the error contours do not change their shapes and sizes 
while the central values of the contours
vary with the cutoff scale $\Lambda$.
The solid-line contours show the analysis which includes the contributions of 
the three-point chiral coefficients in addition to the two point chiral coefficients. 

 Although the contribution of the three-point chiral
coefficients to $\al_1(\Lambda)-\al_0(\Lambda)$  (or $S(\Lambda)-T(\Lambda)$ )
in equations (\ref{rgeal1}) and (\ref{rgeal0}) 
are loop factor suppressed,
the experimental uncertainty of these chiral coefficients 
is almost two orders of magnitude larger than
those of the two-point chiral coefficients, as shown in Table \ref{data}.
Therefore, if we do not make any theoretical assumptions on the magnitude
of the chiral coefficients, the uncertainties of $S(\Lambda)-T(\Lambda)$ 
can be dominated by those of three-point chiral coefficients
at large $\Lambda$. The central values of $S(\Lambda)-T(\Lambda)$ move
according to the central values of the present TGC measurements,
and the size of the solid contours grows with increasing $\Lambda$.

We would like to mention some salient features with respect to Figs. \ref{fig4-j}-\ref{fig6-j}:
{\bf (1)} In the absence of the TGC contribution 
(dashed line  contours),  $S(\Lambda)$ decreases as $\Lambda$ increases 
from the reference LEP1 constraint at $\Lambda = m_Z$. 
This is in agreement with
the observation of Ref. \cite{Bagger, Peskin:2001rw}. 
The dashed contours for $\Lambda=1\textrm{TeV}$ essentially agree
with the $S-T$ contour of Ref. \cite{pdg2006} at $m_H=1 \textrm{TeV}$.

{\bf (2)} When the operators that contribute to the TGC are taken into
account, the $\Lambda$ dependences of the center of the $S-T$ contour
are governed by the central values of the TGC measurements in
Eq. (\ref{tgcdatacase1}), Eq. (\ref{tgcdatacase2}), and Eq. (\ref{tgcdatacase3}) 
for the {\bf case 1} (Fig. \ref{fig4-j}),
the {\bf case 2} (Fig. \ref{fig5-j}), and
the {\bf case 3} (Fig. \ref{fig6-j}), respectively.
For instance, central value of the TGC contribution
to $-\beta_{\al_1}$, $(15 \al_2 - 5 \al_3 + 3 \al_9) g^2/6 $, is $-0.03$
for the {\bf case 1}, Eq. (\ref{tgcdatacase1}),
which adds up to the leading contribution of $-1/12$, and hence the central value
of $S(\Lambda)$ gets even more negative than that of the dashed contours for
$\al_2=\al_3=\al_9=0$. The central value of the same combination is
$0.65$ for the {\bf case 2}, Eq. (\ref{tgcdatacase2}), from the 
L3 measurement \cite{Achard:2004ji}, which makes the central value of $S(\Lambda)$ positive
at large $\Lambda$ in Fig. 5. Large error of this single experiment
constraint explains the rapid growth of the 1 $\sigma$ allowed region with increasing
$\Lambda$. In the {\bf case 3}, where the custodial symmetry violating coefficient
$\al_9$ is allowed to take an arbitrary value, the central values of Eq. (\ref{tgcdatacase3})
make the same combination about $0.08$ which approximately cancels
the constant term of $-1/12$. This results in the almost $\Lambda$-independence
of the central value of $S(\Lambda)$ as observed in Fig. 6.

In summary, those models which satisfy the uncertainty
\bea
(15 \al_2 - 5 \al_3 + 3 \al_9) g^2/6 \approx &
1.09 \al_2 - 0.36 \al_3 + 0.22 \al_9  \nnb \\  > & 1/12
\eea
give $S(\Lambda) \geq S(m_Z)$, and hence QCD-like models can still be consistent
with the electroweak data provided that they give rise to the TGC anomaly which
satisfy the above condition. In case of the $\Lambda$-dependence of the $T$ parameter,
the relevant contribution of the TGC operators in Eq. (\ref{rgeal0}) gives
$0.07 \al_2 + 0.08 \al_3 - 0.05 \al_9$, which tends to be smaller than 
the leading term of $3 {g'}^2/8$. Accordingly,
in all our examples, $T(\Lambda)$ grows with increasing $\Lambda$ as can be
read out from Figs. \ref{fig4-j}-\ref{fig6-j} for the {\bf case 1},
{\bf case 2}, and {\bf case 3}, respectively.

The central values of $S(\Lambda)-T(\Lambda)$ and the errors induced
by TGC measurements can easily be read 
off from Figs. \ref{fig4-j}-\ref{fig6-j}.
We find
\begin{equation}
\begin{array}{rl}
S(1 \,\,\textrm{TeV}) = &-0.15\pm 0.22\,,\\
T(1\,\, \textrm{TeV}) = &+0.47\pm 0.12\,,
\end{array}
\label{stcase1}
\end{equation}
for the {\bf case 1}, Eq. (\ref{tgcdatacase1}), 
with one-parameter fit data and the custodial symmetry constraint.
\begin{equation}
\begin{array}{rl}
S(1\,\, \textrm{TeV}) = &+0.84\pm 0.60\,,\\
T(1 \,\,\textrm{TeV}) = &+0.68\pm 0.18\,,
\end{array}
\label{stcase2}
\end{equation}
for the {\bf case 2}, Eq. (\ref{tgcdatacase2}), 
with two-parameter fit data from L3 measurement \cite{Achard:2004ji} and 
the custodial symmetry constraint, and
\begin{equation}
\begin{array}{rl}
S(1\,\, \textrm{TeV}) = &-0.02\pm 0.20\,,\\
T(1\,\,  \textrm{TeV}) = &+0.52\pm 0.12\,,
\end{array}
\label{stcase3}
\end{equation}
for the {\bf case 3}, Eq. (\ref{tgcdatacase3}),  without
the custodial symmetry constraint.

The central values of the above results can easily
be obtained from the TGC contribution to
the $\beta$ functions of $\al_1$ and $\al_0$
\begin{equation}
\begin{array}{rl}
S(1 \,\,\textrm{TeV}) - S(m_Z)  =& -0.13  
\\ & + 1.66 \al_2  - 0.55 \al_3 + 0.33 \al_9 \,, \\
T(1\,\, \textrm{TeV}) - T(m_Z)  =& 0.40
\\ & + 0.60 \al_2 + 0.70 \al_3 - 0.44 \al_9 \,.
\end{array}
\label{st}
\end{equation}

These numbers and figures demonstrate an apparent fact that
the uncertainty of the TGC measurement 
can significantly affect the value
of $S(\Lambda)-T(\Lambda)$. We point out the fact that 
the current precision of the TGC measurement is
not good enough to fix the sign of $S(\Lambda)$.

\subsection{The uncertainty of $T(\Lambda)$ from the 
three-point and four-point chiral coefficients}

In Figs. (\ref{fig4-j}-\ref{fig6-j}),
the TGC contributions can at most modify the central value of 
$T$(1  TeV) by  
$\left \vert \Delta T(1\,\, {\rm  TeV})\right\vert\approx  0.20$. 
Thus the contribution of TGC is not large enough 
to cancel the large leading contribution 
from $3{g'}^2/8$ in the $\beta$ function of the $\al_0$ (or equally $T$) parameter, 
which makes $T(\Lambda)$ positive.

In this section, we analyze the effect of the four-point chiral coefficients 
to the $T(\Lambda)$ parameter.
Numerical results are given in Table \ref{tableT}. 
The columns $\de T_Z$ and $\de T^{\textrm{TGC}}$
list the $1 \sigma$ uncertainty from the 
measurement of $T$ and the TGC at $\mu=m_Z$, respectively.
The column $\de T^{\textrm{QGC}}$ lists the  
uncertainty from the QGC constrained by the theoretical unitary bounds.
For the TGC constraints, we adopt the {\bf case 3}, or Eq. (\ref{eq:al239})
as an input.

We notice that with the increasing
$\Lambda$ the TGC uncertainty  $\de T^{\textrm{TGC}}$ increases
logarithmically while the QGC uncertainty $\de T^{\textrm{QGC}}$  decreases
rapidly. This is because of the power dependence of the
unitary bounds given in Eq.(\ref{eq:al456710}).
Consequently we find 
that $\de T^{\textrm{QGC}} > \de T^{\textrm{TGC}}$ for $\Lambda < 950\, \textrm{GeV}$
but $\de T^{\textrm{QGC}} < \de T^{\textrm{TGC}}$  for larger $\Lambda$.

\begin{table}[ht]
\begin{center}
\begin{tabular}{|l|  c| c| c| c| }\hline
& & & & \\
 $\Lambda$ & $T(\Lambda)\pm 1\sigma$ &  $\de T_Z$  &  
$\de T^{\textrm{TGC}}$  & $\de T^{\textrm{QGC}}$  \\ 
\hline
& & & &\\
$0.3$ TeV & $0.37\pm 8.91$   &  $\pm 0.14$  & $\pm 0.06$ & $\pm 8.91$ \\
& & & &\\
$0.5$ TeV & $0.41\pm 1.16$   &  $\pm 0.14$  & $\pm 0.08$ & $\pm 1.15$ \\
& & & & \\
$1$ TeV   & $0.52\pm 0.22$   &  $\pm 0.14$  & $\pm 0.12$ & $\pm 0.10$ \\
& & & & \\
$3$ TeV   & $0.73\pm 0.25$   &  $\pm 0.14$  & $\pm 0.17$ & $\pm 0.04$ \\
& & & &\\ \hline
\end{tabular}
\end{center}
\caption{ Values of $T(\Lambda)$ and its uncertainty.
Individual $1 \sigma$ errors from $\de T_Z$, $\de T^{\textrm{TGC}}$ and 
$\de T^{\textrm{QGC}}$ are also shown. }
\label{tableT}
\end{table}
From Table \ref{tableT}, we can conclude that in the constrained EWCL 
parameter space with $1 \sigma$ error in TGC and 
with theoretical unitary bounds on QGC, it is unlikely to have 
a scenario with vanishing $T(\Lambda=1 \textrm{TeV})$ 
while keeping $T(m_Z)=0.09$.

\section{Conclusions and Discussions}
In this work, we study the impacts of all the $11$ dimensionless 
chiral coefficients ($\al_i,\, i=0,\, \cdots,\, 10$)
of the standard electroweak chiral Lagrangian, under the constraints from 
the experimental data
and the perturbative unitarity bounds. We provide the improved RGEs by including
the linear terms of all chiral coefficients. 
By using the improved RGEs, we examine the 
renormalization scale dependence of the oblique parameters 
and update the experimental uncertainty analysis on $S(\Lambda)$ and $T(\Lambda)$. 

We observe that the electroweak precision measurements 
have constrained the oblique parameters, $S$, $T$, and $U$,
and the anomalous TGC.
According to the above experimental uncertainty analysis,
we find that current precision data for the chiral coefficients of the EWCL (as given
in Table \ref{data})
still allow positive $S(\Lambda)$ parameter space, as shown 
in Figs. (\ref{fig4-j}-\ref{fig6-j}), due to the large uncertainty in the
TGC chiral coefficients. Therefore, it is premature to claim that the
sign of $S(\Lambda)$ conclusively rules out the QCD-like EWSB mechanism.
However, the upcoming colliders, with higher sensitivity to the TGC
and QGC, can further reduce the allowed parameter space and help to
pinpoint the correct model of the EWSB.

Before closing, let us briefly discuss the limitations of our analysis.
Most importantly, we have included neither the
two-loop effects of $O(p^2)$ operators 
nor the tree-level contributions of $O(p^6)$ operators.
In the cases when the derivative power counting rule holds, 
the effective couplings
corresponding to the  $O(p^6)$ operators must be suppressed by  $1/(16 \pi^2)^2$, 
in contrast to the $1/(16 \pi^2)$ factor for the $O(p^4)$ operators.
Meanwhile, up to $O(p^6)$ order, the running of $\beta_{\al_i}$'s 
are not affected by the  effective couplings corresponding to the $O(p^6)$
operators. Two loop effects of $O(p^2)$ operators must be suppressed
by the two loop factor, and they must be tiny. Therefore, we expect that
the uncertainties induced by TGC and QGC are dominant when $\Lambda= 1\textrm{TeV}$.

We have restricted our  $\beta$ functions of chiral
coefficients, to retain linear terms and neglect the
terms proportional to $(\al_i g^2)^2$ and $(\al_i {g'}^2)^2$. These terms can be 
be classified as $p^8$ order effects in the derivative power counting rule as depicted in Feynman diagram FIG. (3a). Since the numerical values of $\al_i$ are small, inclusion of  these terms can only render negligible correction to our $S-T$ contours.

The derivative power counting rule in the $\pi$ system has achieved great
phenomenological success and its validity in the EWCL domain is yet to 
be tested by the experiments. From the pure theoretical viewpoint, 
there exist possible theories ({\it e.g.}, large $N_{TC}$ theories
and models with large extra dimensions) 
which can have large deviations in the triple and quartic gauge couplings
from the prediction of the standard model. For example,
in the large $N_{TC}$ theories, we can have the following power counting rules
for the chiral coefficients \cite{Gasser:1984gg}:
\begin{equation}
\begin{array}{rl}
\al_0  \sim & O(N_{TC}^0)\,,\\
\al_i  \sim & O(N_{TC})\,.
\end{array}
\label{eq:largenc}
\end{equation}
In a situation where the two-point chiral coefficients 
$\al_0$, $\al_1$, and $\al_8$ are already tightly constrained, 
one may look for symmetry breaking models  which can have 
large triple and quartic gauge couplings.

\acknowledgments{
We would like to thank Ulrich Parzefall for communication on the TGC
measurements at LEP2, Masaharu Tanabashi and Masayasu Harada 
for stimulating discussions, and Benjamin Grinstein for 
communication on theoretical bounds to chiral coefficients.
This work is supported in part by Grant-in-Aid for Scientific Research 
(\# 18340060) from 
Ministry of Education, Culture, Science and Technology of Japan,
and the JSPS core university programs.
QSY thanks the JSPS fellowship program (\# P03194) and NCTS (Hsinchu, Taiwan) for
financial support. 
The work of KY is supported in part by JSPS Postdoctoral Fellowships for Research Abroad
and the National Science Foundation under Grant No. NSF PHY05-51164.}

\appendix
\def\theequation{\thesection.\arabic{equation}}
\setcounter{equation}{0}
\section{The calculation of the $\beta$ functions of two-point chiral coefficients
\label{app:method}}
We describe the method of calculating  the $\beta$ functions of chiral coefficients 
given in Eqs. (\ref{rgeg}-\ref{rgeal10}) through the following steps \cite{dhy1}.
\begin{enumerate}
\item  In the background field method \cite{Abbott:1981ke}, we decompose 
fields in the Lagrangian of Eq. (\ref{ewl}) as follows:
\bea
W &\rightarrow& {\overline W} + {\widehat W}\,,\\
B &\rightarrow& {\overline B} + {\widehat B}\,,\\
U &\rightarrow& {\overline U} {\widehat U}\,,
\eea
where ${\overline W}$, ${\overline B}$, and ${\overline U}$ are the 
background fields of vector
boson and background Goldstone bosons in the nonlinear realization, 
and ${\widehat W}$, ${\widehat B}$, and ${\widehat U}$ are the
quantum fluctuations.

\item We  parameterize ${\widehat U}$ as
\bea
{\widehat U} & =& \exp\{ \frac{2 {\bf i} \xi^a T^a }{v} \}\,,
\eea
and expand the Lagrangian given in Eq. (\ref{ewl}) up to the quadratic terms  
\bea
{\cal L}_{EW}&=& 
{\cal L}_{EW}({\overline W}, {\overline B}, {\overline U})
\wsep + \frac{\de {\cal L}_{EW}}{\de {\overline W^a}}  \de {\widehat W^a} 
      + \frac{\de {\cal L}_{EW}}{\de {\overline B}} \de {\widehat B}
\wsep + \frac{\de {\cal L}_{EW}}{\de \xi^a} \xi^a
\wsep + \textrm{bilinear terms of ${\widehat W^a}$, ${\widehat B}$, and $\xi^a$}
\wsep + \cdots\,.
\eea

The structure of the divergences upto dimension 4 operators  
induced by the quantum fluctuations are organized as
\bea
\de {\cal L}_{EW}({\overline W}, {\overline B}, {\overline U}) &=&
2 \frac{\de g }{ g} \frac{1}{g^2} {\bar H_1} + 2 \frac{\de g' }{ g'} \frac{1}{{g'}^2} {\bar H_2}
\wsep + \frac{\de v^2}{v^2} v^2 {\bar {\cal L}_{W/Z}}
 + \sbbkf{ \de \al_0 +  \al_0 \frac{\de v^2}{v^2}} v^2 {\bar {\cal L}_0}
\wsep + \sum_{i=1}^{10} \de \alpha_i {\bar {\cal L}^i }\,,
\label{ctt}
\eea
due to the local gauge symmetries of the classical 
fields, divergences of higher order operators are not relevant to our
purpose and are simply thrown away.

\item We cast  the quadratic terms in quantum fields ( the vector bosons ${\widehat V^a}$, 
the Goldstone bosons $\xi^i$, and the ghosts $v^{\bar a}$ and $u^b$) in a
compact form with the appropriate gauge fixing terms as described in 
Appendix \ref{app:ghost} at one loop level.
\begin{equation}
\label{eqn:minimal}
\begin{array}{rl}
{\cal L}_{quad} =
\end{array}
-\frac{1}{2} ({\widehat V}^{\dagger}_{\mu}, \xi^{\dagger} ) \left(\begin{matrix}
\Box_{VV}^{\mu \nu} & \lx^{\mu} \cr 
\rx^{\nu} & \Box_{\xi\xi}^{\prime} \cr
\end{matrix}
\right) 
\left(\begin{matrix}
{\widehat V} \cr 
\xi \cr
\end{matrix}
\right) 
 - {v}^a  \Box_{vu}^{ab} u^b\,.
\end{equation}
Here we have made the partial integrals 
and organized the quadratic terms in the mass eigenstate basis.
Calculation in the weak interaction eigenstate basis yields the same results,
which serves as one extra checking point for our computation method.

\item We express the one-loop  divergences induced by
the quantum fluctuations by using the Gaussian integral over
the $d-$dimensional space-time 
\bea
\int_x {\cal L}_{1-loop}&=& - \frac{1}{ 2}  \left \{ 
   Tr\ln\Box_{V}
+  Tr\ln\Box'_{\xi}
-2 Tr\ln\Box_{vu}
\ssep + Tr\ln\left ( 1 - 
{\stackrel{\rightharpoonup}{X}^{\mu}} \Box^{-1}_{V;\mu\nu} 
{\stackrel{\leftharpoonup}{X}^{\nu}} \Box^{'-1}_{\xi} \right  )
\right \}\,.
\label{log}
\eea

\item Finally by using the heat kernel technique \cite{Vassilevich:2003xt}, 
we extract the divergent terms from  Eq. (\ref{log}) 
which are linear in $\al_i$. The divergences emanating from the first three terms in 
Eq. (\ref{log}) are listed in table \ref{tableV}, \ref{tablex}, and  \ref{tablevu} respectively. 
Tables \ref{tableVxl} and  \ref{tableVxm} lists divergences contributed from the fourth term of  Eq. (\ref{log}).
The  operator $v^2 {\cal L}_0$ gives two terms,
as shown in Eq. (\ref{ctt}). The divergences listed in these tables 
are those of the $\delta \, \al_0 v^2 {\cal L}_0$, 
while those corresponding to $\al_0 \, \de v^2 \,  {\cal L}_0$
are neglected.
\end{enumerate}

\par Combination of all the divergences determines the counter terms given in Eq. (\ref{ctt})
and the RGE given in Eqs. (\ref{rgeg}-\ref{rgev2}). We have used the following equation
of motion to remove the ambiguity of the parameterization of the Goldstone fields
\bea
\partial \cdot {\overline Z} &=& \frac{O^Z_4(\al_i)}{v^2} \,,\\
{\bf d} \cdot  {\overline W^{\pm}}  &=& \pm {\bf i} \frac{\rho g_Z^2 - g^2}{g_Z} {\overline Z} \cdot {\overline W^{\pm} } + \frac{O^W_4(\al_i)}{v^2}\,,\label{eomw}
\eea
where $\rho=1 - 2 \al_0$ and $d \cdot {\overline W^{\pm}} = \partial \cdot {\overline W^{\pm}} \mp {\bf i} e {\overline A} \cdot {\overline W^{\pm}}$.
The terms with $O^{Z,W}_4(\al_i)$ contribute to $O(p^6)$ and higher order 
operators and do not
contribute to the $\beta$ functions of chiral coefficients 
given in Eqs. (\ref{rgeg}-\ref{rgeal10}).

\begin{table*}
\begin{center}
\begin{ruledtabular}
\begin{tabular}{|c|c|c|c|c|c|c|c|c|c|c|c|c|}
                              &                    & $\al_0$ & $\al_1$      & $\al_2$ & $\al_3 $              & $\al_4$ & $\al_5$ & $\al_6$ & $\al_7$ & $\al_8$ & $\al_9$ & $\al_{10}$ \\ \hline
$\frac{1}{g^2}{\bar H}_1$     & $\frac{20}{3}$     &         &              &         & $\frac{40 g^2}{3}$    &         &         &         &         & $4 g^2$ & $ \frac{8 g^2}{3} $ & \\ \hline
$\frac{1}{{g'}^2}{\bar H}_2$  &                    &         &              &         &                       &         &         &         &         &         &                      & \\ \hline
${\bar {\cal L}}_1$                  &                    &         &$-4 g^2$      &$ 4 g^2$ &$-\frac{8 g^2}{3}$     &         &         &         &         &         &                      & \\ \hline
${\bar {\cal L}}_2$                  &                    &         &              &         &                       &         &         &         &         &         &                      & \\ \hline
${\bar {\cal L}}_3$                  &                    &         &              &         &$-\frac{20 g^2}{3}$    &$-2 g^2 $&$4 g^2$  &$-2 g^2$ &$4 g^2$  &         &                      & \\ \hline
${\bar {\cal L}}_4$                  &                    &         &              &         &     &         &         &         &         &         &                      & \\ \hline
${\bar {\cal L}}_5$                  &                    &         &              &         &     &         &         &         &         &         &                      & \\ \hline
${\bar {\cal L}}_6$                  &                    &         &              &         &     &         &         &         &         &         &                      & \\ \hline
${\bar {\cal L}}_7$                  &                    &         &              &         &    &         &         &         &         &         &                      & \\ \hline
${\bar {\cal L}}_8$                  &                    &         &              &         &                       &         &         &         &         &$-12 g^2$& $\frac{16}{3} g^2$  & \\ \hline
${\bar {\cal L}}_9$                  &                    &         &              &         &    &        & &   $2 g^2$      &    $-4 g^2$     &         &$-\frac{20 g^2}{3}$                & \\ \hline
${\bar {\cal L}}_{10}$               &                    &         &              &         &     &         &         &         &         &         &                      & \\ \hline
$v^2 {\bar {\cal L}}_{W/Z}$          &                    &         &$2 g^2 {g'}^2$&$-2 g^2 {g'}^2$ &$-2 g^2 {g'}^2$&$7 g^4 + g_Z^4$ & $10 g^4 + 4 g_Z^4$ &    $g_Z^4$     &   $4 g_Z^4$     & & & \\ \hline
$v^2 {\bar {\cal L}}_0$              &                    &         &$ g^2 {g'}^2$&$- g^2 {g'}^2$ &$- g^2 {g'}^2$& $-\frac{5}{2}{g'}^2 (g^2 + g_Z^2)$ &   $-{g'}^2 (g^2 +g_Z^2)$      &   $-g^4-\frac{11}{2} g_Z^4$     &  $-4 (g^4 + g_Z^4)$       & &  & $-6 g_Z^4$\\
\end{tabular}
\caption{Divergences contributed from $-Tr\ln\Box_{V}/2$ up to linear terms of
$\al_i$. The unit is $1/{8 \pi^2 \epsilon}$. }
\label{tableV}
\end{ruledtabular}
\end{center}
\end{table*}

\begin{table*}
\begin{center}
\begin{ruledtabular}
\begin{tabular}{|c|c|c|c|c|c|c|c|c|c|c|c|c|}
                 &                    & $\al_0$ & $\al_1$ & $\al_2$ & $\al_3 $                        & $\al_4$                  & $\al_5$                      & $\al_6$                  & $\al_7$                  & $\al_8$ & $\al_9$ & $\al_{10}$ \\ \hline
$\frac{1}{g^2}{\bar H}_1$     &$-\frac{1}{12}$ &$\frac{1}{6}$&         &                  &$-\frac{g^2 +g_Z^2}{2}$ &                          &                              &                          &                          & $- g^2$ &                      & \\ \hline
$\frac{1}{{g'}^2}{\bar H}_2$     &$-\frac{1}{12}$ &$\frac{1}{3}$&         &$-g^2$             &                       &                          &                              &                          &                          &         &                      & \\ \hline
${\bar {\cal L}}_1$     &$-\frac{1}{12}$ &              &$g^2$    &$-\frac{g^2}{2}$   &$-\frac{g^2}{2}$        &                          &                              &                          &                          &         & $-\frac{g^2}{2}$    & \\ \hline
${\bar {\cal L}}_2$     &$-\frac{1}{24}$ &$-\frac{1}{6}$&         &$\frac{g^2}{2}$   &                       &$\frac{g^2}{4}$          & $-\frac{g^2}{2}$             &                          &                          &         &                      & \\ \hline
${\bar {\cal L}}_3$     &$-\frac{1}{24}$ & $\frac{1}{6}$&         &                  &$\frac{g^2}{2}$        &$\frac{2 g^2+{g'}^2}{8}$ & $-\frac{2 g^2+{g'}^2}{4}$    &$\frac{2 g^2+{g'}^2}{8}$ & $-\frac{2 g^2+{g'}^2}{4}$&         &   $-\frac{g^2}{2}$   & \\ \hline
${\bar {\cal L}}_4$     & $\frac{1}{12}$ &$1$          &         &                  &                       &$\frac{4 g^2+{g'}^2}{2}$  &     $g^2$                    &$-\frac{g^2-{g'}^2}{2}$   &                          &         &                      & \\ \hline
${\bar {\cal L}}_5$     & $\frac{1}{12}$ &$-\frac{1}{2}$ &         &                  &                       &$-\frac{6 g^2+{g'}^2}{4}$ & $-\frac{4 g^2+{g'}^2}{2}$    &$-\frac{g^2+{g'}^2}{4}$   & $-\frac{3 g^2+{g'}^2}{2}$&         &                      & \\ \hline
${\bar {\cal L}}_6$     &                &$-\frac{3}{2}$ &         &                  &                       &$ \frac{{g'}^2}{4}$       &$ \frac{{g'}^2}{2}$           &$\frac{11g^2+3{g'}^2}{4}$ &   $g^2+{g'}^2$           &         &                      &$-g^2+{g'}^2$   \\ \hline
${\bar {\cal L}}_7$     &                &$ \frac{3}{4}$&         &                  &                       &$-\frac{3 {g'}^2}{8}$     &$-\frac{ {g'}^2}{4}$          &$-\frac{9g^2+7{g'}^2}{8}$ & $-\frac{g^2-3{g'}^2}{4}$ &         &                      &$-2 g^2-{g'}^2$  \\ \hline
${\bar {\cal L}}_8$     &                &$-\frac{1}{2}$ &         &                  &$\frac{{g'}^2}{2}$    &                          &                              &         &         &$3 g^2$  & $-g^2$                & \\ \hline
${\bar {\cal L}}_9$     &                &$-\frac{1}{2}$&         &                  &$\frac{{g'}^2}{2}$     & $-\frac{{g'}^2}{8}$      &    $\frac{{g'}^2}{4}$       &$-\frac{2 g^2+ {g'}^2}{8}$ & $\frac{2 g^2+ {g'}^2}{4}$ &         & $2 g^2$              & \\ \hline
${\bar {\cal L}}_{10}$  &                &              &         &                  &                       &         &         &         &         &         &                      &$ 5 g^2$\\ \hline
$v^2 {\bar {\cal L}}_{W/Z}$ &$\frac{(g^2+g_Z^2)}{4}$ &$\frac{3 }{2} g^2 - g_Z^2 $ &$-\frac{ g^2 {g'}^2 }{2}$& & &$-\frac{7}{4} g^4 -\frac{1}{4} g_Z^4$ & $-\frac{5}{2} g^4 - g_Z^4$ &$-\frac{g_Z^4}{4}$&$-g_Z^4$     &   $\frac{1}{4} g^4$  & & \\ \hline
$v^2 {\bar {\cal L}}_0$     &   $\frac{{g'}^2}{8}$   &$\frac{3 }{2} g^2 -\frac{3}{4} g_Z^2$&$-\frac{ g^2 {g'}^2 }{4}$& & &$\frac{5 {g'}^2 (g^2+g_Z^2)}{8}  $ & $\frac{{g'}^2 (g^2 + g_Z^2)}{4}  $ &$\frac{2}{8} g^4 +\frac{11}{8} g_Z^4$&$g^4+g_Z^4$     &   $\frac{g^4}{8} $  & &$\frac{3 g_Z^4}{2}$\\
\end{tabular}
\caption{Divergences contributed from $-Tr\ln\Box_{\xi}/2$ up to linear terms of
$\al_i$. The unit is $1/{8 \pi^2 \epsilon}$.}
\label{tablex}
\end{ruledtabular}
\end{center}
\end{table*}

\begin{table*}
\begin{center}
\begin{ruledtabular}
\begin{tabular}{|c|c|c|c|c|c|c|c|c|c|c|c|c|}
                 &                    & $\al_0$ & $\al_1$ & $\al_2$ & $\al_3 $              & $\al_4$ & $\al_5$ & $\al_6$ & $\al_7$ & $\al_8$ & $\al_9$ & $\al_{10}$ \\ \hline
$\frac{1}{g^2}{\bar H}_1$     & $\frac{2}{3} $  &         &         &         & $\frac{2 g^2}{3} $   &         &         &         &         &        & $\frac{g^2}{3}$& \\ \hline
$\frac{1}{{g'}^2}{\bar H}_2$     &                    &         &         &         &                       &         &         &         &         &         &                      & \\ \hline
${\bar {\cal L}}_1$     &                    &         &         &         &$-\frac{g^2}{3} $     &         &         &         &         &         &                      & \\ \hline
${\bar {\cal L}}_2$     &                    &         &         &         &                       &         &         &         &         &         &                      & \\ \hline
${\bar {\cal L}}_3$     &                    &         &         &         &$-\frac{g^2}{3} $     &         &         &         &         &         &                      & \\ \hline
${\bar {\cal L}}_4$     &                    &         &         &         &                       &         &         &         &         &         &                      & \\ \hline
${\bar {\cal L}}_5$     &                    &         &         &         &                       &         &         &         &         &         &                      & \\ \hline
${\bar {\cal L}}_6$     &                    &         &         &         &                       &         &         &         &         &         &                      & \\ \hline
${\bar {\cal L}}_7$     &                    &         &         &         &                       &         &         &         &         &         &                      & \\ \hline
${\bar {\cal L}}_8$     &                    &         &         &         &                       &         &         &         &         &         & $-\frac{g^2}{3}$ & \\ \hline
${\bar {\cal L}}_9$     &                    &         &         &         &                       &         &         &         &         &         & $-\frac{g^2}{3}$ & \\ \hline
${\bar {\cal L}}_{10}$     &                    &         &         &         &                       &         &         &         &         &         &  & \\ \hline
$v^2 {\bar {\cal L}}_{W/Z}$ &                    &         &$-2 g^2 {g'}^2$&$g^2 {g'}^2$ &$g^2 {g'}^2$                      &   &        &       &     &    & & \\ \hline
$v^2 {\bar {\cal L}}_0$     &                    &         &$-  g^2 {g'}^2$&$\frac{1}{2} g^2 {g'}^2 $ &$\frac{1}{2} g^2 {g'}^2  $&  &         &       &      & &  & \\
\end{tabular}
\caption{Divergences contributed from $Tr\ln\Box_{vu}$ up to linear terms of
$\al_i$. The unit is $1/{8 \pi^2 \epsilon}$. }
\label{tablevu}
\end{ruledtabular}
\end{center}
\end{table*}

\begin{table*}
\begin{center}
\begin{ruledtabular}
\begin{tabular}{|c|c|c|c|c|c|c|c|c|c|c|}
                 & $\al_1$           & $\al_2$                            & $\al_3 $                           & $\al_4$                     & $\al_5$           & $\al_6$                    & $\al_7$                   & $\al_8$            & $\al_9$              & $\al_{10}$ \\ \hline
$\frac{1}{g^2}{\bar H}_1$     &   ${g'}^2$        &$-\frac{5 {g'}^2}{6}$                &$-\frac{11g^2}{3}+{g'}^2$            &                             &                   &                            &                           & $g^2$              & $-\frac{5 g^2}{6} $   & \\ \hline
$\frac{1}{{g'}^2}{\bar H}_2$     &   $2 g^2$         &$  2 g^2$                          &$-\frac{5g^2}{3}$                    &                             &                   &                            &                           &                    &                      & \\ \hline
${\bar {\cal L}}_1$     &   $-g^2$          &$-g^2$                               &$ \frac{8g^2}{3}$                  &                             &                   &                            &                           & $g^2$              &       $g^2$         & \\ \hline
${\bar {\cal L}}_2$     &$\frac{g^2}{2}$   &$-\frac{5g^2}{2}-\frac{{g'}^2}{12}$ &$-\frac{5g^2}{12}$                  &$-\frac{g^2+{g'}^2}{2}$      &$g^2-{g'}^2$      &$-\frac{{g'}^2}{2}$          &$-{g'}^2$                   &                    &$\frac{g^2}{12}$      &   \\ \hline
${\bar {\cal L}}_3$     &$\frac{{g'}^2}{4}$&$-\frac{{g'}^2}{6}$                 &$\frac{19 g^2}{12}-\frac{{g'}^2}{4}$&$ \frac{g^2-{g'}^2}{2}$      &$-g^2$             &$\frac{g^2-{g'}^2}{2}$     &$-g^2$                      &$-\frac{g^2}{4}$     &$\frac{13g^2}{6}$     &   \\ \hline
${\bar {\cal L}}_4$     &                   &$\frac{5{g'}^2}{6}$                &$\frac{g^2}{2}$                    &$-\frac{15 g^2+13{g'}^2}{2}$ &$-3g^2-3{g'}^2$    &$g^2-4{g'}^2$               &$-2{g'}^2$                 &                    &$\frac{g^2}{6}$      &   \\ \hline
${\bar {\cal L}}_5$     &                   &$-\frac{{g'}^2}{3}$                 &$\frac{g^2}{2}$                    &$ \frac{5 g^2+8 {g'}^2}{2}$  &$g^2+2{g'}^2$      &$\frac{g^2+2 {g'}^2}{2}$    &$3g^2-{g'}^2$              &                    &$\frac{g^2}{3}$      &   \\ \hline
${\bar {\cal L}}_6$     &                   &$-{g'}^2$                            &                                    &$ 6 {g'}^2$                  &$3 {g'}^2$         &$-\frac{18 g^2-{g'}^2}{2}$  &$-3g^2-{g'}^2$             &                    &                      &$2g^2-6{g'}^2$\\ \hline
${\bar {\cal L}}_7$     &                   &$\frac{{3g'}^2}{4}$                &                                    &$-\frac{19 {g'}^2}{4}$       &$-\frac{5{g'}^2}{2}$&$\frac{7 g^2-13 {g'}^2}{4}$&$-\frac{7 g^2+2 {g'}^2}{2}$&                    & $-\frac{g^2}{4}$      &$4g^2-3{g'}^2$\\ \hline
${\bar {\cal L}}_8$     &   $-{g'}^2$       &$\frac{5 {g'}^2}{6}$               &$-{g'}^2$                            &                             &                   &                            &                           &$-3 g^2$            & $-\frac{7 g^2}{6}$    & \\ \hline
${\bar {\cal L}}_9$     &$\frac{{g'}^2}{4}$&$\frac{{g'}^2}{12}$                 &$-\frac{7{g'}^2}{4}$                &${g'}^2$                    &${g'}^2$          &$-\frac{g^2-2 {g'}^2}{2}$    &$g^2+{g'}^2$              &$\frac{3{g'}^2}{4}$& $-\frac{59 g^2}{12}$ &   \\ \hline
${\bar {\cal L}}_{10}$  &                   &                                    &                                    &                             &                   &$\frac{17{g'}^2}{4}$        &$\frac{7{g'}^2}{2}$        &                    &                      &$-14g^2+9{g'}^2$\\ 
\end{tabular}
\caption{Divergences (Part A) contributed from $- Tr\ln \left ( 1 - 
{\stackrel{\rightharpoonup}{X}^{\mu}} \Box^{-1}_{V;\mu\nu} 
{\stackrel{\leftharpoonup}{X}^{\nu}} \Box^{'-1}_{\xi} \right  )/2 $ up to 
linear terms of $\al_i$. The unit is
$1/{8 \pi^2 \epsilon}$. }
\label{tableVxl}
\end{ruledtabular}
\end{center}
\end{table*}

\begin{table*}
\begin{center}
\begin{ruledtabular}
\begin{tabular}{|c|c|c|c|c|c|c|c|}
                     &                    & $\al_0$             & $\al_1$           & $\al_2$                               & $\al_3 $            & $\al_8$          & $\al_9$              \\ \hline
$v^2 {\bar {\cal L}}_{W/Z}$ &  $-2 g^2 - {g'}^2$ & $-2 g^2 + 4 {g'}^2$  &$2 g^2 {g'}^2$     &$-2 g^2 {g'}^2 + \frac{3}{2} {g'}^4 $   &$-2 g^2 {g'}^2+2 g^4$&$- g^4$           & $ g^4 - \frac{3g^2 {g'}^2}{2}  $ \\ \hline
$v^2 {\bar {\cal L}}_0$     & $-\frac{ {g'}^2}{2}$&$-3 g^2 + 3{g'}^2$   &$g^2 {g'}^2$      &$- g^2 {g'}^2   + \frac{3}{4} {g'}^4 $ &$- g^2 {g'}^2$       & $-\frac{g^4}{2}$&$\frac{ g^4}{2}- \frac{3g^2 {g'}^2}{4}  $ \\
\end{tabular}
\caption{Divergences (Part B) contributed from $- Tr\ln \left ( 1 - 
{\stackrel{\rightharpoonup}{X}^{\mu}} \Box^{-1}_{V;\mu\nu} 
{\stackrel{\leftharpoonup}{X}^{\nu}} \Box^{'-1}_{\xi} \right  )/2 $ up to 
linear terms of $\al_i$. The unit is
$1/{8 \pi^2 \epsilon}$. }
\label{tableVxm}
\end{ruledtabular}
\end{center}
\end{table*}

\setcounter{equation}{0}
\section{Gauge fixing terms and ghost terms
\label{app:ghost}}
In order to cast the quantum fluctuations of vector 
boson fields into the minimal form of Eq. (\ref{eqn:minimal}),
we make a special choice of the  gauge fixing parameters.
In the background field method, the covariant gauge fixing terms
are given as 
\bwt
\bea
{\cal L}_{GF,A}&=&-\frac{G_A }{ 2} \sbbkf{\partial\cdot\hA + f_{AZ} \partial\cdot\hZ  - i f_{AW} (\hWm \cdot \bWp - \hWp \cdot \bWm)}^2\cma
\label{gfa}
\eea
\bea
{\cal L}_{GF,Z}&=&-\frac{G_Z }{ 2} \sbbkf{\partial\cdot\hZ + f_{Z\xi} \xi_Z  + i f_{ZW} (\hWm \cdot \bWp - \hWp \cdot \bWm)}^2\cma
\label{gfz}
\eea
\bea
{\cal L}_{GF,W}&=&- G_W \sbbkf{d\cdot\hWp + f_{W\xi} \xi_W^+
        + i f_{WZ} \bZ \cdot \hWp - i p_{WZ} \bWp \cdot \hZ
        + i p_{WA} \bWp \cdot \hA}\nnb\\&&\sbbkf{d\cdot\hWm + f_{W\xi} \xi_W^-
        - i f_{WZ} \bZ \cdot \hWm + i p_{WZ} \bWm \cdot \hZ
        - i p_{WA} \bWm \cdot \hA}\,.
\label{gfw}
\eea
\ewt
These gauge fixing terms explicitly guarantee the $U_{\textrm{EM}}^{}(1)$
symmetry. Gauge fixing parameters can be uniquely determined as follows 
by requiring that the quadratic terms to have the minimal compact form given in
Eq. (\ref{eqn:minimal}): 
\begin{enumerate}
\item The gauge parameters related with kinetic terms of the
propagators of vector bosons are given as:
\bea
 G_A &=& C_1 \,,\eb
 G_Z &=& C_3 - \frac{C_2^2}{C_1} \,,\eb
 G_W &=& 1\,.
\label{gfpara1}
\eea
\item  The gauge parameter related with the kinetic mixing between
the quantum fields $\hA$ and $\hZ$ is given as
\bea
 f_{AZ} &=&\frac{C_2}{C_1^2} \,.
\label{gfpara2}
\eea

\item  The gauge parameters related with the couplings between
$\hA$ ($\hZ$) and $\hWp(\hWm)$ are given as
\bea
 f_{AW} &=& \frac{e}{G_A}\,,\eb
 f_{ZW} &=& \frac{1}{G_Z} \frac{C_7}{ 2} + \frac{e }{ G_Z}  \frac{C_2 }{ C_1}\,,\eb
 p_{WA} &=& \frac{C_5 }{ 2}\,,\eb
 p_{WZ} &=& \frac{C_6 }{ 2}\,,\eb
 f_{WZ} &=& \frac{C_7 }{ 2}\,.
\label{gfpara3}
\eea
These gauge parameters guarantee that the covariant differential operator
of vector sector has an Hermitian form. 

\item  The gauge parameters related with the mixing between the quantum vector
bosons and the Goldstone bosons are determined as
\bea
 f_{Z\xi} &=& \frac{\rho}{G_Z} \frac{g_Z v}{2}\,,\eb
 f_{W\xi} &=& - \frac{1}{G_W}\frac{g v}{2}\,.
\label{gfpara4}
\eea
The parameter $f_{Z\xi}$ takes into account the diagonalization and the normalization
of the $Z$ vector boson and the $\xi_Z$ Goldstone boson.
\end{enumerate}
Here, the parameters $C_i$  are defined as
\bea
 C_1 &=& 1 - \frac{g^2 g^{'2}}{g_Z^2} \left( 2 \alpha_1 + \alpha_8 \right)\,,\label{c2aa}\eb
 C_2 &=& \frac{g g^{'}}{g_Z^2} \left[ \alpha_1 \left( g^2 - g^{'2}\right)  + \alpha_8 g^2 \right ]\,,\eb
 C_3 &=& 1 + \frac{g^2}{g_Z^2} \left(  2 \alpha_1 g^{'2}  - \alpha_8 g^2 \right)\,,\eb
 C_5 &=& 2\frac{ g g^{'}}{g_Z} \left[ 1 - (\alpha_1 + \alpha_8  
\right. \nnb \\ && \left. -\alpha_2 - \alpha_3 -
     \alpha_9 ) g^2 \right ] ,\eb
 C_6 &=& 2\frac{ g^2}{g_Z} \left[ 1 -  (\alpha_8 - \alpha_3  - \alpha_9 )g^2 \right. \nnb \\ && \left. + (\alpha_1
    - \alpha_2) g^{'2} \right]\,,\eb
 C_7 &=& 2 \frac{g^2 }{g_Z} \left( 1 + \alpha_3 g_Z^2\right) .
\eea

From the gauge fixing terms given in Eqs. (\ref{gfa}-\ref{gfw}), the determinant of
the ghost terms are found as
\bea
\det{\left( \frac{\de F }{ \de \al} \right ) } &=& \det \left ( {D' \cdot D} + \frac{\de F }{ \de \xi} \frac{\de \xi }{ \de \al} \right )\,,
\eea
where $i D'_{\mu}$ is non-Hermitian with
\bea
D'_{\mu} = C_{gh} \partial_{\mu} + \Gamma'_{\mu}\,,
\eea
where the matrix $C_{gh}$ is 
\begin{equation}
\mathbf{C_{gh}} =\left (
\begin{array}{cccc}
\sqrt{G_{A}} & \sqrt{G_A} \, f_{AZ} & 0 &0 \\
0 & \sqrt{G_Z} & 0 &0 \\
0 & 0 & 1 &0 \\
0 & 0 & 0 &1 
\end{array}
\right )\,,
\end{equation}
and the matrix $\Gamma'$ is 
\bwt
\begin{equation}
\mathbf{\Gamma'} =\left (
\begin{array}{cccc}
0 & 0 & i \sqrt{G_{A}} f_{AW} {\bar W^{-}} & - i \sqrt{G_{A}} f_{AW} {\bar W^{+}} \\
0 & 0 & - i \sqrt{G_Z} f_{ZW} {\bar W^{-}} &  i \sqrt{G_Z} f_{ZW} {\bar W^{+}} \\
i p_{WA} {\bar W^{+}} &- i p_{WZ} {\bar W^{+}} & -i e {\bar A} + i f_{WZ} {\bar Z} &0 \\
- i p_{WA} {\bar W^{-}} & i p_{WZ} {\bar W^{-}} & 0 & i e {\bar A} - i f_{WZ} {\bar Z} 
\end{array}
\right )\,.
\end{equation}
\ewt
Here we observe that Hermiticity of this determinant is broken 
by both the kinetic parameter $C_{gh}$ and the gauge potential
terms $\Gamma'$ by the chiral coefficients $\al_i$.  
The parameter $C_{gh}$ is determined 
by the chiral coefficients $\al_{1}$ and $\al_8$, which induce 
the mixing between photon and
the Z boson. In $\Gamma'$, Hermiticity is broken by
the chiral coefficients $\al_2$, $\al_3$, and $\al_9$.
On the other hand, $ D= \partial + \Gamma$ is Hermitian and $\Gamma$ is given as
\bwt
\begin{equation}
\left ( \begin{array}{cccc} 
0 & 0 & i e {\bar W^{-}} & - i e {\bar W^{+}} \\
0 & 0 &-i \frac{g^2 }{ g_Z} {\bar W^{-}} & i \frac{g^2 }{ g_Z} {\bar W^{+}} \\
 i e {\bar W^{+}} &-i \frac{g^2}{ g_Z} {\bar W^{+}} & -i e {\bar A} + i \frac{g^2 }{ g_Z} {\bar Z} &0 \\
-i e {\bar W^{-}} & i \frac{g^2}{ g_Z} {\bar W^{-}} & 0 & i e {\bar A} - i \frac{g^2 }{ g_Z} {\bar Z} 
\end{array}
\right )\,.
\end{equation}
\ewt

The term $(\de F/ \de \xi) (\de \xi/\de \al)$ is 
\begin{equation}
\left ( \begin{array}{cccc} 
0 & 0 & 0 & 0 \\
0 & - \frac{\rho}{\sqrt{G_Z}} \frac{g_Z^2 v^2}{4}  & 0 & 0 \\
0 & 0 & - \frac{g^2 v^2}{4} &0 \\
0 & 0 & 0 & - \frac{g^2 v^2}{4} 
\end{array}
\right )\,.
\end{equation}

We can make the ghost determinant to be Hermitian, by using the fact that 
\bea
\det(\frac{\de F^a}{ \de \al^{\bar b}})&=& \frac{ \det( M^{ab} \frac{\de F^b}{\de \al^{\bar a}} {\overline M^{{\bar a}{\bar b}}})}
{\det(M^{ab}) \det({\overline M^{{\bar a}{\bar b}}})}\,.
\eea
This identity is justified  
if  $\det M^{ab}$ and $\det {\overline M^{{\bar a} {\bar b}}}$ do
not vanish or go to infinity. In the ghost term, the
degree of freedom to choose the matrices $M^{ab}$ 
and ${\overline M^{{\bar a} {\bar b}}}$ reflects
the fact that there are two types of real ghost fields we can introduce,
which are labeled as $v$ type and $u$ type ghosts, respectively.

By properly choosing the matrices $M^{ab}$ and ${\overline M^{{\bar a}{\bar b}}}$,
we can reorganize the ghost determinant as
\bea
\det( M^{ab} \frac{\de F^b}{\de \al^{\bar a}} {\overline M^{{\bar a}{\bar b}}})
 &=&  D_{gh} \cdot D_{gh} \wsep + \sigma_{gh}^{mass} + \sigma_{gh}^{2}\,,
\eea
where $D_{gh}$ is Hermitian. The mass matrix $\sigma_{gh}^{mass}$ is 
the same as that of the vector bosons. There are several ways to adjust
the matrices $M^{ab}$ and ${\overline M^{{\bar a}{\bar b}}}$ in order
to make the Feynman rules of the ghost sector well-defined. 
Although we cannot make the term
$\sigma_{gh}^{2}$ Hermitian, up to linear terms of $\al_i$, we find 
our results given in Table \ref{tablevu} is independent of the procedures
to hermitize the ghost determinant.


\begin{thebibliography}{10}

\bibitem{ws}
        S. Weinberg, Phys. \ Rev. \ {\bf D 13} (1976) 974; Phys. \ Rev. \ {\bf D 19} (1979) 1277;
        L. Susskind,  Phys. \ Rev. \ {\bf D 20} (1979) 2619;

\bibitem{hill}
        C. T. Hill and E. H. Simmons,
  Phys.\ Repts.\ {\bf 381} (2003) 235 [ Erratum-ibid. {\bf 390} (2004) 553].

\bibitem{our}
        S. Dutta, K. Hagiwara, Q.S. Yan, unpublished, arXiv:hep-ph/0603038;
        Q.S. Yan, arXiv:hep-ph/0703189, talk presented at SCGT2006 workshop, 
        Nagoya, Japan (November 2006).

\bibitem{Gasser:1983yg}
  J.~Gasser and H.~Leutwyler,
  Annals Phys.\  {\bf 158} (1984) 142;

\bibitem{Gasser:1984gg}
  J.~Gasser and H.~Leutwyler,
  Nucl.\ Phys.\ B {\bf 250} (1985) 465.

\bibitem{Harada:2003jx}
  M.~Harada and K.~Yamawaki,
  Phys.\ Rept.\  {\bf 381} (2003) 1
  [arXiv:hep-ph/0302103].

\bibitem{Appelquist:1980ae}
  T.~Appelquist and C.~W.~Bernard,
  Phys.\ Rev.\ D {\bf 22}, 200 (1980);
  Phys.\ Rev.\ D {\bf 23}, 425 (1981).

\bibitem{Longhitano:1980iz}
  A.~C.~Longhitano,
  Phys.\ Rev.\ D {\bf 22}, 1166 (1980).

\bibitem{Longhitano:1980tm}
  A.~C.~Longhitano,
  Nucl.\ Phys.\ B {\bf 188}, 118 (1981).

\bibitem{Appelquist:1993ka}
  T.~Appelquist and G.~H.~Wu,
  Phys.\ Rev.\ D {\bf 48}, 3235 (1993)
  [arXiv:hep-ph/9304240].

\bibitem{Bagger}
J.~A.~Bagger, A.~F.~Falk and M.~Swartz,
Phys.\ Rev.\ Lett.\  {\bf 84}, 1385 (2000)
[hep-ph/9908327].

\bibitem{Heister:2001qt}
  A.~Heister {\it et al.}  [ALEPH Collaboration],
  Eur.\ Phys.\ J.\ C {\bf 21}, 423 (2001)
  [arXiv:hep-ex/0104034].

\bibitem{Abbiendi:2003mk}
  G.~Abbiendi {\it et al.}  [OPAL Collaboration],
  Eur.\ Phys.\ J.\ C {\bf 33}, 463 (2004)
  [arXiv:hep-ex/0308067].

\bibitem{Achard:2004ji}
  P.~Achard {\it et al.}  [L3 Collaboration],
  Phys.\ Lett.\ B {\bf 586}, 151 (2004)
  [arXiv:hep-ex/0402036].

\bibitem{Schael:2004tq}
  S.~Schael {\it et al.}  [ALEPH Collaboration],
  Phys.\ Lett.\ B {\bf 614}, 7 (2005).

\bibitem{Abazov:2005ys}
  V.~M.~Abazov {\it et al.}  [D0 Collaboration],
  arXiv:hep-ex/0504019.

\bibitem{georgi}
        H. Georgi, Annu. Rev. Nucl. Part. Sci. {\bf 43} (1993) 209.

\bibitem{Herrero:1993nc}
  M.~J.~Herrero and E.~Ruiz Morales,
  Nucl.\ Phys.\ B {\bf 418}, 431 (1994)
  [arXiv:hep-ph/9308276];
  Nucl.\ Phys.\ B {\bf 437}, 319 (1995)
  [arXiv:hep-ph/9411207];
  S.~Dittmaier and C.~Grosse-Knetter,
  Nucl.\ Phys.\ B {\bf 459}, 497 (1996)
  [arXiv:hep-ph/9505266].

\bibitem{Barbieri:2004qk}
  R.~Barbieri {\it et. al. },
  Nucl.\ Phys.\ B {\bf 703}, 127 (2004).




\bibitem{SSVZ}
P.~Sikivie, L.~Susskind, M.~Voloshin and V.~Zakharov,
Nucl.\ Phys.\ {\bf B173}, 189 (1980).


\bibitem{pandt}
M.~E.~Peskin and T.~Takeuchi,
Phys.\ Rev.\ Lett.\  {\bf 65}, 964 (1990);
Phys.\ Rev.\  {\bf D46}, 381 (1992).

\bibitem{Peskin:2001rw}
  M.~E.~Peskin and J.~D.~Wells,
  Phys.\ Rev.\ D {\bf 64}, 093003 (2001).

\bibitem{pdg2006}
  W.~M.~Yao {\it et al.}  [Particle Data Group],
  J.\ Phys.\ G {\bf 33}, 1 (2006).
  
\bibitem{unknown:2005em}
    [ALEPH Collaboration],
  Phys.\ Rept.\  {\bf 427}, 257 (2006)
  [arXiv:hep-ex/0509008].

\bibitem{Hagiwara:1998}
  K.~Hagiwara, S.~Matsumoto, D.~Haidt and C.~S.~Kim,
  Z.\ Phys.\ C {\bf 64}, 559 (1994)
  [Erratum-ibid.\ C {\bf 68}, 352 (1995)]
  [arXiv:hep-ph/9409380];
  K.~Hagiwara, D.~Haidt and S.~Matsumoto,
  Eur.\ Phys.\ J.\ C {\bf 2}, 95 (1998)
  [arXiv:hep-ph/9706331].
  K. Hagiwara,
Annu.\ Rev. \ Nucl.\ Part. \ Sci. {\bf 48} 463, (1998).

\bibitem{Arbuzov:2005ma}
 D.~Y.~Bardin, P.~Christova, M.~Jack, L.~Kalinovskaya, A.~Olchevski, S.~Riemann and T.~Riemann,
  Comput.\ Phys.\ Commun.\  {\bf 133}, 229 (2001)
  [arXiv:hep-ph/9908433]; 
  A.~B.~Arbuzov {\it et al.},
  Comput.\ Phys.\ Commun.\  {\bf 174}, 728 (2006)
  [arXiv:hep-ph/0507146].
  
   
\bibitem{topmass}
  E.~Brubaker {\it et al.}  [Tevatron Electroweak Working Group],
  arXiv:hep-ex/0608032.


\bibitem{Hagiwara:1986vm}
  K.~Hagiwara, R.~D.~Peccei, D.~Zeppenfeld and K.~Hikasa,
  Nucl.\ Phys.\ B {\bf 282}, 253 (1987).

\bibitem{Baur:1987mt}
  U.~Baur and D.~Zeppenfeld,
  Phys.\ Lett.\ B {\bf 201}, 383 (1988).

\bibitem{LEPEWWG}
          LEPEWWG/TGC/2005-01, WWW access at http://www.cern.ch/LEPEWWG/lepww/tgc.

\bibitem{Cornwall:1973tb}
  J.~M.~Cornwall, D.~N.~Levin and G.~Tiktopoulos,
  Phys.\ Rev.\ Lett.\  {\bf 30}, 1268 (1973)
  [Erratum-ibid.\  {\bf 31}, 572 (1973)].
  J.~M.~Cornwall, D.~N.~Levin and G.~Tiktopoulos,
  Phys.\ Rev.\ D {\bf 10}, 1145 (1974)
  [Erratum-ibid.\ D {\bf 11}, 972 (1975)].
  C.~H.~Llewellyn Smith,
  Phys.\ Lett.\ B {\bf 46}, 233 (1973).
  S.~D.~Joglekar,
  Annals Phys.\  {\bf 83}, 427 (1974).
  B.~W.~Lee, C.~Quigg and H.~B.~Thacker,
  Phys.\ Rev.\ D {\bf 16}, 1519 (1977).
  B.~W.~Lee, C.~Quigg and H.~B.~Thacker,
  Phys.\ Rev.\ Lett.\  {\bf 38}, 883 (1977).

\bibitem{Distler:2006if}
  J.~Distler, B.~Grinstein, R.~A.~Porto and I.~Z.~Rothstein,
  Phys.\ Rev.\ Lett.\  {\bf 98}, 041601 (2007)
  [arXiv:hep-ph/0604255].
  
\bibitem{dhy1}
  Q.~S.~Yan and D.~S.~Du,
 Phys.\ Rev.\ D {\bf 69}, 085006 (2004);
  S. Dutta, K. Hagiwara, and Q. ~S, ~Yan, Nuc. Phys. {\bf B 704}, 75 (2005).

\bibitem{private}
        M. Tanabashi, private communication.

\bibitem{Chivukula:2007ic}
  R.~S.~Chivukula, S.~Matsuzaki, E.~H.~Simmons and M.~Tanabashi,
  arXiv:hep-ph/0702218.




\bibitem{Abbott:1981ke}
  L.~F.~Abbott,
  Acta Phys.\ Polon.\ B {\bf 13} (1982) 33.

\bibitem{Vassilevich:2003xt}
  D.~V.~Vassilevich,
  Phys.\ Rept.\  {\bf 388}, 279 (2003).

\end{thebibliography}
\end{document}